\documentclass[preprint,12pt, 3p]{elsarticle}

\usepackage{amsmath}
\usepackage{amssymb}
\usepackage{mathrsfs}
\usepackage{hyperref}
\usepackage{bm}
\usepackage{accents}
\usepackage{wrapfig}

\newdefinition{rmk}{Remark}

\journal{arXiv}

\begin{document}

\begin{frontmatter}

\title{A chain stretch-based gradient-enhanced model for damage and fracture in elastomers}

\author[address1]{S. Mohammad Mousavi}
\author[address2,address3]{Jason Mulderrig}
\author[address4]{Brandon Talamini}
\author[address1,address5,address6]{Nikolaos Bouklas\corref{corr}}
\ead{nb589@cornell.edu}
\address[address1]{Sibley School of Mechanical and Aerospace Engineering, \\ Cornell University, Ithaca, NY 14853, USA}
\address[address2]{Materials and Manufacturing Directorate, Air Force Research Laboratory, Wright-Patterson Air Force Base, Dayton, OH 45433, USA}
\address[address3]{National Research Council (NRC) Research Associateship Programs, The National Academies of Sciences, Engineering, and Medicine, 500 Fifth St., N.W., Washington, DC, 20001, USA}
\address[address4]{Lawrence Livermore National Laboratory, Livermore, CA 94550, USA}
\address[address5]{Center for Applied Mathematics, Cornell University, NY 14853, USA}
\address[address6]{Pasteur Labs, Brooklyn, NY 11205, USA}



\cortext[corr]{Corresponding authors}


\begin{abstract}
Similar to other quasi-brittle materials, it has been recently shown that elastomers can exhibit a macroscopically diffuse damage zone that accompanies the fracture process. 
In this study, we introduce a novel stretch-based gradient-enhanced damage (GED) model that allows the fracture to localize and also captures the development of a physically diffuse damage zone. This capability contrasts with the paradigm of the phase field method for fracture, where a sharp crack is numerically approximated in a diffuse manner. Capturing fracture localization and diffuse damage in our approach is achieved by considering nonlocal effects that encompass network topology, heterogeneity, and imperfections. These considerations motivate the use of a statistical damage function dependent upon the nonlocal deformation state. From this model, fracture toughness is realized as an output. While GED models have been classically utilized for damage modeling of structural engineering materials (e.g., concrete), they face challenges when trying to capture the cascade from damage to fracture, often leading to damage zone broadening (de Borst and Verhoosel, 2016). This deficiency contributed to the popularity of the phase-field method over the GED model for elastomers and other quasi-brittle materials. Other groups have proceeded with damage-based GED formulations that prove identical to the phase-field method (Lorentz \textit{et al.}, 2012), but these inherit the aforementioned limitations. To address this issue in a thermodynamically consistent framework, we implement two modeling features (a nonlocal driving force bound and a simple relaxation function) specifically designed to capture the evolution of a physically meaningful damage field and the simultaneous localization of fracture, thereby overcoming a longstanding obstacle in the development of these nonlocal strain- or stretch-based approaches. We discuss several numerical examples to understand the features of the approach at the limit of incompressibility, and compare them to the phase-field method as a benchmark for the macroscopic response and fracture energy predictions. 

\end{abstract}


\end{frontmatter}

\section{Introduction}\label{Section:Intro}

Elastomers and hydrogels are composed of flexible polymer chains cross-linked together to form a network structure. Due to this microscopic network topology,  these materials are able to endure large deformations while exhibiting substantial toughness, which makes them well-suited for load-bearing applications \cite{drury2003hydrogels, nonoyama2016double}. These remarkable properties have inspired extensive theoretical research and advancements in understanding their behavior \cite{bouklas2012swelling, bouklas2015effect,kang2010variational, bouklas2015nonlinear}. These materials play a crucial role in conventional engineering applications and have recently gained attention as potential candidates for next-generation soft robotics and biomedical devices \citep{gent2012engineering, zhalmuratova2020reinforced}. Despite their desirable mechanical properties, they are vulnerable to damage and fracture under high loads, which can weaken their structural integrity and limit their applicability \citep{mark2003elastomers, sun2012highly, itskov2016rubber, tehrani2017effect, bai2017fatigue}. Notably, elastomeric materials exhibit flaw sensitivity. For flaw sensitive elastomers, small cracks and defects often do not critically affect mechanical performance, but larger flaws exceeding a material-specific fractocohesive length scale significantly reduce stretch-to-rupture capacity \citep{chen2017flaw, zhou2021flaw, yang2019polyacrylamide}. To elaborate more, elastomer fracture occurs when polymer chains break under applied loads; as the material undergoes increasing deformation, the polymer chains stretch, causing the bonds within the chain backbones and crosslinks to elongate. Eventually, these bonds reach their breaking point (which is loading rate-dependent \citep{mulderrig2023statistical, yang2020multiscale}), ultimately leading to fracture. As recently observed, elastomer fracture can be accompanied by diffuse chain breakages that accumulate in the vicinity of the crack tip and its wake \citep{slootman2020quantifying}. This process has been observed to have signatures that transcend the network mesh size and are relevant to continuum size scales, often reported on the order of the fractocohesive length scale (up to $mm$). There are also many works that have discussed the diffuse nature of chain rupture due to network level chain load-sharing and network imperfections \citep{verron2017equal, mulderrig2021affine,li2020variational,lamont2021rate}. It is essential to gain a more profound understanding of the processes involved in damage and fracture within soft materials to enhance advancements in the design of polymer networks for industrial and cutting-edge engineering applications. Additionally, it is crucial to develop numerical approaches that can probe the transition from damage to fracture in elastomer networks, capturing phenomena such as flaw sensitivity to reliably predict the failure cascade in these materials systems.

Recent experimental research has shed light on the intricate transition from damage to fracture in elastomers \cite{zhou2021flaw, yang2019polyacrylamide, slootman2020quantifying, lin2020fracture, lin2021fracture, zheng2022fracture, lei2022network, ducrot2014toughening, morelle20213d, sanoja2021mechanical, slootman2022molecular, ju2024role}, while theoretical and computational methods continue to evolve to address these challenges. Traditional macroscopic phenomenological models often fail to accurately represent the microstructural complexities of these materials \cite{xiang2018general, chen2020mechanically}, limiting their predictive capability, particularly in relation to damage and fracture. Additionally, constraints such as incompressibility further increase the difficulty of modeling these phenomena. The Lake-Thomas \citep{lake1967strength} fracture criterion often falls short of providing accurate quantitative predictions for fracture toughness in elastomers. As such, several works have attempted to further enhance the conceptual understanding of fracture in elastomers as well as the robustness of predictive capabilities beyond the simple scenario that the Lake-Thomas theory portrays \cite{wang2019quantitative, wang2023contribution, beech2023reactivity, deng2023nonlocal, wang2024loop, hartquist2025scaling, hartquist2025fracture, arora2020fracture, arora2021coarse, arora2024effect, wang2024fracture, fan2024discontinuous, wang2024fresh, persson2024influence, yu2025shortest, barney2022fracture, zhang2024predicting, li2020elongation}. Even though there has been significant progress in the field, the need for advanced damage modeling frameworks is crucial to better predict mechanical failure in elastomers while accounting for all mechanisms at play, especially towards accounting for diffuse damage mechanisms. 

Over the past few decades, several numerical methods have been developed to model fracture. These approaches generally fall into two categories based on the way they represent sharp cracks. In the discrete modeling approach, fracture surfaces are realized by modifying the geometry of the original, intact structure incorporating a sharp crack boundary \cite{de2016gradient}. Common techniques in this category include remeshing \cite{CAMACHO19962899, Ingraffea1985}, the extended finite element method (XFEM) \cite{moes1999finite, sukumar2000extended, khoei2023modeling}, and cohesive zone modeling \citep{cazes2009thermodynamic, cuvilliez2012finite}. Although they are appealing due to the explicit representation of cracks as geometrical discontinuities, they encounter significant challenges: achieving robust implementation for handling complex discontinuities can become particularly difficult, especially in the three-dimensional setting \cite{de2016gradient}. To avoid this problem, smeared modeling approaches were introduced. The core feature of the smeared modeling approach is that sharp crack discontinuities are numerically approximated over a finite width through the introduction of a damage field. The damage field $d$ inclusively assumes a scalar value between 0 and 1 ($d\in[0,1]$), where 0 indicates no damage and 1 indicates a fully damaged material. 

Moving from damage to fracture, the smeared modeling approach can be further classified into two categories: local and nonlocal continuum damage models. In the context of elastomers, various local continuum damage models have been developed. These models vary based on the representative volume element (RVE) employed for homogenization, such as the chain conformation space \citep{lamont2021rate, buche2021chain}, the eight-chain model \citep{lei2021recent, mao2017rupture, xiao2021modeling, zhao2021multiscale, lu2020pseudo}, and the full-network microsphere model \citep{mulderrig2021affine, arunachala2021energy, dal2009micro, guo2021micromechanics}. However, despite their progress, local continuum models experience a loss of well-posedness when expressed in the context of boundary value problems, resulting in mesh-dependent finite element solutions. This dependency pertains not only to the level of mesh refinement but also to the orientation of the mesh, which raises concerns about its physical validity \cite{de2016gradient}. In order to circumvent mesh dependency while also representing physically meaningful phenomena at the microscale and network scale, nonlocal continuum damage models have been developed. These models have become popular among the computational mechanics community. Two of the most utilized approaches to nonlocal continuum damage modeling are the phase-field method and gradient-enhanced damage (GED) models. 

In the context of small strain and linear elasticity, phase field fracture has seen significant improvements through various contributions. These include improved algorithmic implementations \cite{miehe2010rate}, the inclusion of dynamic effects \cite{borden2012phase}, more efficient hybrid formulations that lower computational costs \cite{ambati2015review}, new models that factor in plastic strain-dependent fracture toughness \cite{yin2020ductile}, and recent advancements in crack nucleation \citep{kumar2018fracture, kumar2020phase, talamini2021attaining}. Together, these developments enhance both the accuracy and efficiency of phase-field simulations in fracture mechanics. When applying the phase-field method to soft materials like polymers and elastomers, it is essential to account for large deformations and near incompressibility. Additionally, if possible, the method should incorporate lower-scale information from polymer chain statistical mechanics and network topology characteristics. At higher Poisson ratios, a Lagrange multiplier may be added to enforce the incompressibility constraint, but it presents numerical challenges due to difficulties in satisfying the inf-sup condition \cite{vassilevski1996preconditioning, benzi2005numerical, loghin2004analysis}. To overcome these issues, various numerical strategies have been proposed, such as using Taylor-Hood spaces (which are above the lowest order) \cite{taylor2000mixed, onate2004finitecalculus, gavagnin2020stabilized, mang2020phase, alessi2020phase, suh2020phase, cajuhi2018phase, wriggers2021taylor} and other stabilization techniques \cite{klaas1999stabilized, maniatty2002higher, ang2022stabilized}. Building on these computational and theoretical advances, several recent studies have successfully applied the phase-field method to elastomer fracture situations \cite{li2020variational, swamynathan2022phase, arunachala2023multiscale, feng2023phase, ye2023nonlinear, zhao2023phase, pranavi2024unifying}. 

As one focuses on network and chain level features of elastomers, damage, and fracture can be described from the perspective of statistical mechanics of polymer chains, providing a rich foundation to continuum theories. Elastomer fracture is a hierarchical process where the rupture of individual polymer chains at the microscale leads to macroscopic material failure. To obtain multiscale information and predictive capabilities for the damage-to-fracture transition, it is essential to capture the behavior of polymer chains up to their rupture point. Polymer chain rupture is driven primarily by bond stretching, which is an enthalpic rather than an entropic process, as noted by Lake and Thomas (1967) \cite{lake1967strength} and Wang \textit{et al.} (2019) \cite{wang2019quantitative}. Early models, such as the Kuhn and Gr{\"u}n (1942) \cite{kuhn1942beziehungen} freely-jointed chain (FJC) model, were phenomenologically extended by Smith \textit{et al.} (1996) \cite{smith1996overstretching} and Mao \textit{et al.} (2017) \cite{mao2017rupture} to account for bond extensibility. Mao \textit{et al.} (2017) \cite{mao2017rupture} specifically incorporated the effect of bond extensibility into the Helmholtz free energy function to improve predictions of chain rupture. Recent works by Buche \textit{et al.} (2021, 2022) \cite{buche2021chain, buche2022freely} introduced the $u$FJC model, which incorporates arbitrary bond potentials into the FJC framework. Although these advances represent a significant step forward, the complexity of the $u$FJC model limits its direct application in continuum-scale simulations. To address this, Mulderrig \textit{et al.} (2023) \cite{mulderrig2023statistical} developed a simplified ``composite'' $u$FJC model that retains the essential features of bond extensibility while remaining analytically tractable. That work additionally focused on the statistics of polymer chain rupture, identifying an maximal upper bound for the stretch a chain can experience while allowing for thermally activated chain scission to occur at lower levels of chain stretch.
 
At the network level, elastomers exhibit considerable structural heterogeneity due to network topology imperfections and chain-length polydispersity, which strongly influence their mechanical response under deformation. Capturing the effects of this underlying heterogeneity is critical for understanding elastomer fracture, and assumptions such as equal strain \cite{itskov2016rubber, tehrani2017effect, mulderrig2021affine, diani2019fully} or equal force load-sharing \cite{verron2017equal, mulderrig2021affine, li2020variational} among chains provide different ways to model how the network responds to stress. Additionally, establishing a connection between chain-level deformation and continuum-level response has been accomplished through several micro-to-macro homogenization techniques. These techniques include the affine three-chain model \citep{wang1952statistical}, the non-affine four-chain model \citep{flory1943statistical}, the non-affine Arruda-Boyce eight-chain model \citep{arruda1993three}, the affine full-network microsphere model \citep{treloar1979non, wu1992improved, wu1993improved}, and several non-affine full-network microsphere models \citep{mulderrig2021affine, arunachala2021energy, guo2021micromechanics, diani2019fully, miehe2004micro, tkachuk2012maximal, ghaderi2020physics, rastak2018non, govindjee2019fully, xiao2021micromechanical, zhan2023new}. By combining these chain- and network-level representations, continuum models that incorporate bond extensibility-driven damage provide a bottom-up framework for modeling elastomer fracture. In this type of approach, damage and fracture should manifest from chain-level information, and not the other way around. Therefore, contrary to the phase-field method, fracture toughness should not be an input, but rather, an output of the approach. In this context, gradient-enhanced damage models could serve as a suitable nonlocal continuum damage tool that allows for the integration of information from statistical mechanics while accounting for network-level heterogeneity and physically diffuse damage representations. These models start from the definition of a nonlocal variable (often equivalent strain or damage) that is representative of the behavior of a region-of-influence of the continuum \cite{peerlings1996gradient}. This nonlocal continuum region-of-influence depends on underlying micromechanics characteristics and necessitates the introduction of a relevant length scale to the model.

The primary applications of GED models are for subsurface modeling, concrete, and metals \cite{pham2010gradient, kuhl2000anisotropic, marigo2016overview, seupel2019gradient, zhao2023modified, saji2024new}. Historically, nonlocal strain-based and damage-based GED models have been extensively used in the literature for damage and fracture simulations \citep{peerlings1996gradient,peerlings1998gradient}, but underlying limitations of the approach as discussed in de Borst and Verhoosel (2016) \citep{de2016gradient} limited their predictive capabilities in fracture problems due to issues such as fictitious damage zone broadening. In the past decade, the works of Lorentz (2011, 2012, 2017) \cite{lorentz2011gradient,lorentz2012modelling,lorentz2017nonlocal} and Talamini \textit{et al.} (2018) \cite{talamini2018progressive} have provided important steps towards the appropriate implementation of damage-based GED models for fracture. In the past few years, GED models have gained popularity for analysis of damage and failure in soft materials. Specifically, Talamini \textit{et al.} (2018) \cite{talamini2018progressive} developed a theory for modeling progressive failure in elastomeric materials, focusing on the microscopic physics of bond scission rather than the traditional macroscopic energy release rate theory. Their proposed model distinguishes between the entropic contributions to free energy arising from the configurational entropy of polymer chains and the internal energy contributions resulting from bond deformation. Additionally, Valverde-González \textit{et al.} (2023) \cite{valverde2023locking} introduced two novel gradient-enhanced continuum damage formulations, Q1Q1E24 and Q1Q1P0, designed to tackle shear and volumetric locking issues in compressible and nearly incompressible hyperelastic materials. Lamm \textit{et al.} (2024) \cite{lamm2024gradient1} introduced a comprehensive framework for modeling crack propagation in polymers under large deformations using gradient damage models. Their work addressed the limitations of local damage models by incorporating a global damage variable based on the micromorphic balance equation \cite{lamm2024gradient1, forest2009micromorphic}. They further advanced this approach by enhancing the fully thermomechanically coupled model to incorporate viscoelastic effects \cite{lamm2024gradient2}.

As polymer chain statistical mechanics models are described in terms of chain stretch, it is most natural for GED formulations that try to capture this type of information to be stretch-based as well. But to date, there have been no successful implementations of strain-based or stretch-based GED that can accurately describe fracture (for any class of materials). Mousavi \textit{et al.} (2024) \cite{mousavi2024evaluating}, after discussing the effect of artificial viscosity in phase-field fracture, proposed a stretch-based GED model that eliminates the need for fracture toughness as a model input. However, their study highlighted the persisting issue of damage zone broadening even when strain/stretch-based GED is used for elastomers, as earlier identified by de Borst and Verhoosel (2016) \citep{de2016gradient}. This leaves a significant gap in the literature, as there is currently no strain/stretch-based GED formulation that can consistently overcome damage zone broadening, with the possible exception of some recent works by Wosatko and co-authors \citep{wosatko2021comparison,wosatko2022survey} focusing on concrete modeling. This gap is crucial, as filling it would be the first step towards a natural upscaling of polymer chain statistical mechanics information to the continuum through a GED formulation. Such a GED formulation would be intrinsically based upon a nonlocal chain stretch that encompasses information about network topology, imperfections, and polydispersity.

This work aims to address this gap by developing a nonlocal chain stretch-based gradient-enhanced damage model for fracture in near-incompressible hyperelastic materials. Even though this effort is motivated by statistical mechanics, it is purposely restricted to the continuum level to overcome all numerical issues associated with a convergent strain/stretch-based GED. Damage zone broadening is a serious problem hindering engineers from utilizing GED in fracture modeling across different classes of materials. In this work, we address this problem by first building upon previous elastomer GED and phase-field modeling efforts. We then develop the necessary modeling features to prevent the emergence of damage-zone broadening while allowing for a physically diffuse representation of damage that strongly interacts with the initiation and propagation of cracks. More intriguingly, these features are developed by considering the statistical mechanics underpinnings of polymer chain rupture and eventually allow the calculation of fracture toughness as a simulation output. 
The outline of the paper is as follows: In Section \ref{Section:theory}, we present the nonlocal chain stretch-based GED model following the principle of virtual power. Section \ref{Section:NumericalImplementation} describes the numerical implementation carried out using the open-source finite element platform \texttt{FEniCS} \cite{alnaes2015fenics}. Section \ref{Section:results} consists of five subsections: first, we apply the traditional GED model to illustrate the broadening issue. Then, in the following subsections, we incorporate and validate our modifications to the traditional GED formulations to eliminate the broadening issue. Next, we validate our proposed approach by comparing the results with those obtained from the phase-field method, which is representative of experimental findings but also serves as a validation of our fracture energy predictions. The corresponding implementation in \texttt{FEniCS} is available on GitHub\footnote{\url{https://github.com/MMousavi98/Improved_GED}}.
\section{Formulation}\label{Section:theory}
\begin{figure}[h!]
    \centering
    \includegraphics[width=0.6\linewidth]{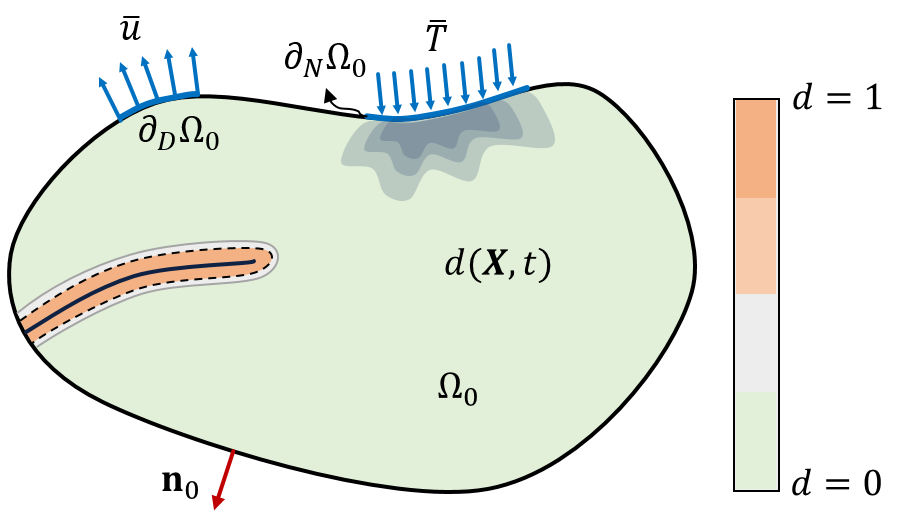}
    \caption{Schematic of a solid continuum that has undergone diffuse damage and fracture.}
    \label{fig:potato}
\end{figure}

In all formulations and modeling processes, it is assumed that ${\Omega}_0 \subset \mathbb{R}^3$ is an open, bounded, and connected region with a sufficiently smooth boundary, labeled as $\partial \Omega_0$. The solid continuum is shown in Fig. \ref{fig:potato}, along with boundary conditions, the sharp crack discontinuity, and physically diffuse damage surrounding the crack discontinuity. In Fig. \ref{fig:potato}, the Dirichlet and Neumann boundary conditions are applied on $\partial_D\Omega_0$ and $\partial_N\Omega_0$, respectively (where $\partial_D\Omega_0\subset {\Omega}_0$ and $\partial_N\Omega_0\subset {\Omega}_0$).

\subsection{Kinematics}\label{SubSection:Kinematics}

As the body undergoes deformation, its points move from their original positions $\bm{X}$ in the reference configuration $\Omega_0$ to new positions $\bm{x}$ in the current configuration $\Omega$. This motion is expressed by the displacement vector $\bm{u}$, where $\bm{u} = \bm{x} - \bm{X}$. This vector describes the shift of a particle from its referential position to its current position. However, to thoroughly describe the deformation, it is necessary to investigate the kinematics of infinitesimal line elements, which can be explained through the following equation,
\begin{equation}\label{bulk-element}
d\bm{x}=\textbf{F}(\bm{X},t)d\bm{X},
\end{equation}
where $\textbf{F}$ is the deformation gradient, defined as 
\begin{equation}\label{deformation-gradient}
\textbf{F}(\bm{X},t)=\frac{\partial\bm{x}}{\partial\bm{X}}.
\end{equation}
The Jacobian determinant of the deformation gradient, expressed as $J(\bm{X}, t) = dv/dV = \det\textbf{F}(\bm{X}, t) > 0$, represents the volume ratio between a current and a referential volume element. Furthermore, the right Cauchy-Green deformation tensor $\textbf{C}$ is introduced as
\begin{equation}
\textbf{C}=\textbf{F}^T\textbf{F},
\end{equation}
along with its principal invariants \cite{holzapfel2002nonlinear},
\begin{equation} \label{invariants}
\begin{split}
    I_1 =\textrm{tr}(\textbf{C}), \\
    I_2 =\frac{1}{2}((\textrm{tr}(\textbf{C}))^2 - \textrm{tr}(\textbf{C}^2)), \\
    I_3= \textrm{det}(\textbf{C}).
\end{split}
\end{equation}

To characterize the behavior of polymer chains in elastomeric materials, a scalar field known as the chain stretch is defined as $\lambda_{ch}={r}/{r_0}$, where $r_0$ is the equilibrium length of a chain and $r$ is the final length of the chain upon deformation. 
By using the 8-chain model proposed by Arruda and Boyce \cite{Arruda-Boyce1993}, the average chain stretch at each material point in the elastomer can be expressed as follows (assuming uniformity in the number of segments per chain),
\begin{equation}\label{Lambda_chain_ch}
\lambda_{ch}=\sqrt{\frac{I_1}{3}}.
\end{equation}

\subsection{Thermodynamic considerations}

In our previous work \citep{mousavi2024evaluating}, we briefly introduced a GED model for elastomers, but identified several issues with the approach. To rectify these issues, we start here with a more detailed derivation of the framework. To fully characterize a thermodynamic system, we first need to identify all relevant state variables. For near-incompressible hyperelastic materials, displacement $(\mathbf{u})$ and hydrostatic pressure $(p)$ are primary state variables that describe the system. However, when modeling damage and fracture, additional state variables are necessary to capture the full behavior of the system. 

It is important to recognize that a polymer network, unlike an ordered crystal, can be characterized by numerous heterogeneities and imperfections that contribute to chain-level load-sharing as well as long-range force transmission, and must be considered in modeling the damage-to-fracture cascade. To incorporate these effects, we introduce a nonlocal chain stretch $\Bar{\lambda}$ as another state variable. Finally, to consider the local deterioration of material integrity, we introduce a scalar damage field $d$. This state variable incorporates information from the nonlocal (chain) stretch $\Bar{\lambda}$ through a damage function that could further take into account thermally activated chain scission, network topology, and other effects (as will be further discussed). With these state variables defined, we can employ the principle of virtual power to establish the governing equations.

The internal mechanical power $P_{int}$ for a polymer network in the reference configuration $\Omega_0$ is expressed as
\begin{equation}\label{eq::rateIntPower}
    P_{int} = \int_{\Omega_0}\left[\mathbf{P}\colon\nabla\dot{\bm{u}} + f_{p}\dot{p} + f_{\Bar{\lambda}}\dot{\Bar{\lambda}} + \boldsymbol{\xi}_{\Bar{\lambda}}\cdot\nabla\dot{\Bar{\lambda}} + f_{d}\dot{d}\right]dV,
\end{equation}
where $\mathbf{P}$ is the first Piola-Kirchhoff stress tensor and, $f_{p}$, $f_{\Bar{\lambda}}$, $\boldsymbol{\xi}_{\Bar{\lambda}}$, and $f_{d}$ represent power conjugates to the corresponding internal state variables, and $dV$ denotes an infinitesimal volume element of $\Omega_0$. By neglecting body forces, the external mechanical work $W_{ext}$ can be expressed as follows,
\begin{equation}
    W_{ext} = \int_{\partial_N\Omega_0}\bm{T}\cdot\bm{u}dA + \int_{\partial_N\Omega_0}\hat{\iota}\Bar{\lambda} dA.
\end{equation}
Here, $\bm{T}$ is the mechanical surface traction and $\hat{\iota}$ defines micromorphic tractions. The external mechanical power $P_{ext}$ is simply
\begin{equation}
    P_{ext} = \dot{W}_{ext} = \int_{\partial_N\Omega_0}{\bm{T}}\cdot\dot{\bm{u}} dA + \int_{\partial_N\Omega_0}\hat{\iota}\dot{\Bar{\lambda}} dA.
\end{equation}
By satisfying the principle of virtual power (first law of thermodynamics in variational form, $\delta{P}_{int} = \delta{P}_{ext}$) and then applying the divergence theorem, the following governing equations and boundary conditions are obtained:
\begin{itemize}
    \item Mechanical equilibrium equation and boundary conditions: \begin{equation} \label{mechanical_equilibrium_governing}
    \begin{split}
        \nabla\cdot\mathbf{P}= \mathbf{0} \quad \text{in} \quad \Omega_0,\\
        \bm{u}= \Bar{\bm{u}} \quad \text{on} \quad \partial_{D}{\Omega}_0,\\
        \mathbf{P}\cdot\bm{n}_0 = \bm{T} \quad \text{on} \quad \partial_N\Omega_0.
        \end{split}
    \end{equation}
    \item {Power conjugate of pressure:}
    \label{pressure_governing}
    \begin{equation}
        f_{p} = 0.
    \end{equation}
    \item Micro-force equilibrium equation and boundary conditions: \begin{equation} \label{chain_governing}
    \begin{split}
        \nabla\cdot\boldsymbol{\xi}_{\Bar{\lambda}}-f_{\Bar{\lambda}} = 0 \quad \text{in} \quad \Omega_0,\\
        \Bar{\lambda}= \Bar{\Bar{\lambda}} \quad \text{on} \quad \partial_{D}{\Omega}_0,\\
        \qquad\boldsymbol{\xi}_{\Bar{\lambda}}\cdot\bm{n}_0 = \hat{\iota} \quad \text{on} \quad \partial_{N}{\Omega}_0.
        \end{split}
    \end{equation}
    \item {Power conjugate of damage:}
    \begin{equation} \label{damage_governing1}
        f_{d} = 0.
    \end{equation}
\end{itemize}

The Helmholtz free energy is considered to be a function of the external and internal variables $\Psi(\mathbf{F},p, \Bar{\lambda}, \nabla\Bar{\lambda}, d)$. Following this specification, the rate of Helmholtz free energy can be determined using the chain rule,
\begin{equation}\label{eq::rateFreeEnergy}
    \frac{\textnormal{d}\Psi}{\textnormal{d} t} = \frac{\partial\Psi}{\partial\mathbf{F}}\colon\dot{\mathbf{F}} + \frac{\partial \Psi}{\partial p}\dot{p} + \frac{\partial \Psi}{\partial \Bar{\lambda}}\dot{\Bar{\lambda}} + \frac{\partial \Psi}{\partial \nabla\Bar{\lambda}}\cdot\nabla\dot{\Bar{\lambda}} + \frac{\partial \Psi}{\partial d}\dot{d}.
\end{equation}
Following the Coleman-Noll procedure to satisfy the Clausius-Planck dissipation inequality, we combine Eqs. \eqref{eq::rateIntPower} and \eqref{eq::rateFreeEnergy} for the rate of dissipation as
\begin{equation}\label{eq::rateDis}
   \mathcal{D} = (\mathbf{P}-\frac{\partial\Psi}{\partial\mathbf{F}})\colon\dot{\mathbf{F}} + (f_p-\frac{\partial \Psi}{\partial p})\dot{p} + (\boldsymbol{\xi}_{\Bar{\lambda}}-\frac{\partial \Psi}{\partial \Bar{\lambda}})\dot{\Bar{\lambda}} + (f_{\Bar{\lambda}}-\frac{\partial \Psi}{\partial \nabla\Bar{\lambda}})\cdot\nabla\dot{\Bar{\lambda}} + (f_d-\frac{\partial \Psi}{\partial d})\dot{d} \geq 0\,.
\end{equation}
For the first four contributions the following constitutive relations are obtained,
\begin{equation}\label{eq::constitutive}
    \mathbf{P} = \frac{\partial\Psi}{\partial\mathbf{F}} ,\quad
    f_p=\frac{\partial \Psi}{\partial p} ,\quad
    \boldsymbol{\xi}_{\Bar{\lambda}} = \frac{\partial \Psi}{\partial \nabla\Bar{\lambda}} ,\quad
    f_{\Bar{\lambda}} = \frac{\partial \Psi}{\partial \Bar{\lambda}}  \quad \text{in} \quad \Omega_0.
\end{equation}

Lastly, in order to satisfy the Clausius-Planck dissipation inequality \eqref{eq::rateDis} for the fifth term and noting that $f_d=0$, according to Eq. \eqref{damage_governing1},  the following must hold true :
\begin{equation} \label{eq:reduced-cp-dissipation-inequality-version-1}
\frac{\partial\Psi}{\partial d}\dot{d} \leq 0~\text{in}~\Omega_0.
\end{equation}
Considering damage as a irreversible process, without loss of generality, the damage state variable $d$ can be defined as monotonically non-decreasing with time; as such, its rate is non-negative, $\dot{d}\geq 0$. Following, to satisfy the dissipation inequality one should satisfy
\begin{equation} \label{eq:reduced-cp-dissipation-inequality-version-2}
\frac{\partial\Psi}{\partial d}\leq 0~\text{in}~\Omega_0.
\end{equation}
which can be satisfied based on further constitutive choices.

\subsection{Model specialization}
\subsubsection{Helmholtz free energy}

The first objective of this section is the specialization of the Helmholtz free energy density function. Prior to focusing on the nonlocal damage problem, we focus on enforcing the near-incompressibility constraint that is often encountered for elastomeric materials. For a local description of an elastomeric material, the Helmholtz free energy density can be preliminarily denoted as $\psi(\mathbf{F})$. To impose the near-incompressibility constraint, a perturbed Lagrangian approach \citep{wriggers2006computational} can be applied following our previous formulations for phase-field fracture in elastomers at the limit of incompressibility \cite{li2020variational, ang2022stabilized}. This method stems from the pure Lagrange multiplier formulation for enforcing exact incompressibility, and then proceeds by assuming a constitutive law for a pressure-like field $p$, enforcing this via an additional Lagrange multiplier, and finding the stationary point from the respective Euler-Lagrange equation. This results in the following free energy density function,
\begin{equation}\label{PF_strain_energy2}
\widehat{\psi}(\mathbf{F},p) = \psi(\mathbf{F}) -p\left(J -1\right) -\frac{p^2}{2\kappa},
\end{equation}
where $\kappa$ is a variable that regulates compressibility. 

To account for the elastic contribution of polymer chains that are allowed to rupture in an irreversible process, we assume an additive decomposition $\Psi=\Psi^{\text{loc}}+\Psi^{\text{nloc}}$ to consider local and nonlocal contributions to the free energy density. The local component aims to represent the elastic energy as a function of deformation, pressure, and damage. This term is a degraded form of the free energy function in Eq. \eqref{PF_strain_energy2},  as
\begin{equation}\label{psi_loc}
    \Psi^{loc}(\mathbf{F},p,d) = a(d)\psi(\mathbf{F}) -b(d)p\left(J -1\right) -\frac{p^2}{2\kappa}\,.
\end{equation}
where $a(d)$ and $b(d)$ are degradation functions.

Building on the aforementioned previous studies \citep{li2020variational,ang2022stabilized}, it is important to note that $b$ has a higher $d$-polynomial order than $a$. This guarantees that the effective resistance to bulk deformation decreases more rapidly than the resistance to shear as driven by the evolution of the damage field. In this way, the volume of the continuum is allowed to increase due to crack opening while the undamaged continuum remains firmly at the incompressibility limit \citep{ang2022stabilized}. This arrangement ensures that the incompressibility constraint does not hinder the physical opening of cracks. For the purposes of this paper, we select $\psi(\mathbf{F})$ as the Neo-Hookean free energy density function, acknowledging its limitations as they pertain to the accurate description of polymer chains at their scission limit. It can be expressed as
\begin{equation}\label{neo-hookean}
{\psi}(\mathbf{F}) =\frac{\mu}{2}(I_1(\mathbf{C})-3-2\ln J),
\end{equation}
where $\mu$ is the shear modulus. A future task for the authors will be to select a $\psi(\mathbf{F})$ function derived from (extensible) polymer chain statistical mechanics considerations following Mulderrig {\textit{et al.}} (2023) \citep{mulderrig2023statistical}.

Meanwhile, the nonlocal component of the free energy, $\Psi^{nloc}$, considers the energetic contributions resulting from the network-level response and its interaction with the macroscopic continuum. The formulation of $\Psi^{nloc}$  is based on the works of \cite{poh2017localizing} and \cite{sarkar2020thermo}, unless stated otherwise. An additive decomposition is postulated as follows, $ \Psi^{nloc} = \Psi_{micmac}^{nloc} + \Psi_{grd}^{nloc}$, where
$\Psi_{micmac}^{nloc}$ considers the deviation between the nonlocal chain stretch $\Bar{\lambda}$, which is defined based on microscopic/network-level considerations, and the local chain stretch  $\lambda_{ch}$, which is defined based on the macroscopic deformation only. This leads to a micro-macro energy penalty:
\begin{equation} \label{psi_nloc_micmac}
    \Psi_{micmac}^{nloc}\left(\lambda_{ch}, \Bar{{\lambda}}\right) = \frac{h_{nl}}{2}\left[\lambda_{ch} - \Bar{\lambda}\right]^2,
\end{equation}
where $h_{nl}$ controls the intensity of the micro-macro nonlocal interactions, and its numerical value is implied to be non-zero and positive ($h_{nl}>0$).
The $\Psi_{grd}^{nloc}$ term accounts for the contributions of the nonlocal chain stretch field via its gradient $\nabla\Bar{\lambda}$ (defined in referential coordinates) as
\begin{equation} \label{psi_nloc_grd}
    \Psi_{grd}^{nloc}\left(\nabla\Bar{\lambda},d\right) = \frac{h_{nl}g(d)\ell^2}{2} \nabla\Bar{\lambda}\cdot\nabla\Bar{\lambda}.
\end{equation}
where $g(d)$ is a relaxation function and will be discussed further in an upcoming Section.

\subsubsection{Degradation functions}
The specific model form is chosen in line with our previous works in phase-field modeling \citep{li2020variational,ang2022stabilized} and GED \citep{mousavi2024evaluating}, where $a(d)$ and $b(d)$ are degradation function that is defined as
\begin{equation}\label{ab}
\begin{split}
a(d) &= (1-k_{\ell})(1-d)^2+k_{\ell}, \\
b(d) &= (1-k_{\ell})(1-d)^3+k_{\ell}.
\end{split}
\end{equation}
The numerical conditioning parameter $k_{\ell}$ is introduced for numerical stability. 

\subsubsection{Relaxation function} \label{subsection:relaxation_function}
\begin{figure}[h!]
    \centering
    \includegraphics[width=0.6\linewidth]{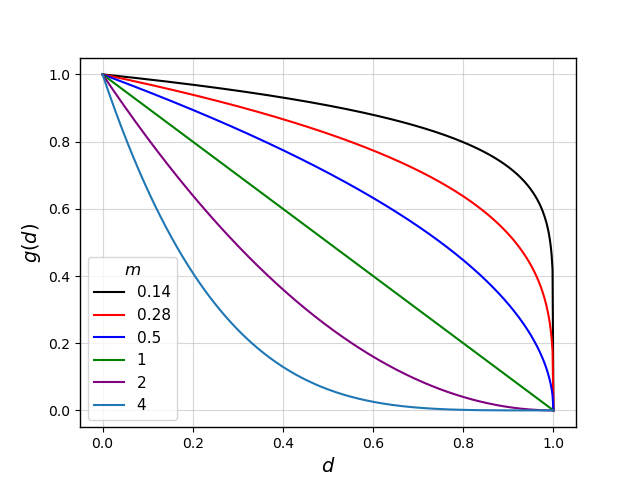}
    \caption{The effect of different relaxation function parameters $m$ on the relaxation function $g(d)$.}
    \label{fig:g_plot}
\end{figure}

Motivated from previous works \cite{wosatko2021comparison, wosatko2022survey}, the relaxation function $g(d)$  diminishes the micro-macro nonlocal interactions in the damage zone. Physically, this can be motivated by focusing on the limit of complete deterioration; information in the continuum is not expected to be transferred (in a nonlocal fashion) from completely damaged regions to undamaged or moderately damaged regions. This relaxation function dictates the progressive deterioration of nonlocal information passing from regions of higher damage to regions of lower damage. We will thoroughly analyze the impact of this term on the simulation in Section \ref{Section:results}. Additionally, $\ell$ is a length scale that arises, defining the size of the nonlocal interaction domain and accounts for the aforementioned network-level effects on the damage and fracture processes. In this way, $\ell$ can be properly interpreted as an intrinsic material parameter. Note that in the phase-field method, the nonlocal term is also multiplied by a length scale. However, in the context of the phase-field method, this length scale has no physical interpretation -- it stems from the regularization approach and is purely a numerical artifact. Motivated from the previous relaxation functions proposed in the literature \cite{wosatko2021comparison, geers1998strain, saroukhani2013simplified}, we use the following simple function in our study:
\begin{equation} \label{g}
    g(d) = (1-d)^m,
\end{equation}
where $m>0$ is the relaxation function parameter and determines the effectiveness of the function. Fig. \ref{fig:g_plot} shows this effect for different values of $m$. 
As $m$ decreases, the interaction shows a sharp decline at higher levels of damage, indicating that this modeling feature will have minimal impact on the problem until the damage becomes significant in the region. As we will see in Section \ref{Section:results}, this behavior of small $m$ values prevents the simulation from developing a high residual stress in the damage zone.

\subsubsection{Constitutive relations}
Following Eqs. \eqref{psi_loc}, \eqref{psi_nloc_micmac}, and \eqref{psi_nloc_grd}, the Helmholtz free energy is given by
\begin{align} \label{eq::Helmholtz}
    \Psi(\mathbf{F},p, \Bar{\lambda}, \nabla\Bar{\lambda}, d) = &~a(d)\psi(\mathbf{F}) -b(d)p\left(J -1\right) -\frac{p^2}{2\kappa} \nonumber \\
    & + \frac{h_{nl}}{2}\left[\lambda_{ch} - \Bar{\lambda}\right]^2 + \frac{h_{nl}g(d)\ell^2}{2} \nabla\Bar{\lambda}\cdot\nabla\Bar{\lambda}.
\end{align}
and satisfies the requirements from Eq. \ref{eq:reduced-cp-dissipation-inequality-version-2}. Combining Eqs. \ref{eq::constitutive} and \ref{eq::Helmholtz}, the following constitutive relations are obtained:
\begin{equation}\label{constitutives}
\begin{split}
    \mathbf{P} = a(d)\frac{\partial\psi}{\partial\mathbf{F}} - b(d) p J \mathbf{F}^{-T} \quad \text{in} \quad \Omega_0, \\
    f_p=-b(d)(J-1)-\frac{p}{k} \quad \text{in} \quad \Omega_0, \\
    \boldsymbol{\xi}_{\Bar{\lambda}} = h_{nl}g(d)\ell^2\nabla\Bar{\lambda}\quad \text{in} \quad \Omega_0, \\
    f_{\Bar{\lambda}} =  -h_{nl}\left[\lambda_{ch} - \Bar{\lambda}\right] \quad \text{in} \quad \Omega_0.
\end{split}
\end{equation}

\subsection{Damage function}

In strain/stretch-based GED models, the damage function must be defined in terms of the nonlocal variable. For rate-dependent formulations, the rate of the primary nonlocal variable can be involved in that definition as well. In this thermodynamically consistent derivation, we start by defining a constitutive law for the thermodynamic conjugate force of the damage as
\begin{equation}\label{damage_conjugateforce}
f_d(d,\Bar{\lambda}) = 
\begin{cases} 
d & \text{if } \Bar{\lambda} < \lambda_{cr} \\
d-1 +\frac{\lambda_{cr}-1}{\mathcal{H}(\Bar{\lambda})-1}\left(1 - c + c e^{-\gamma(\mathcal{H}(\Bar{\lambda}) - \lambda_{cr})}\right) & \text{if } \Bar{\lambda} \geq \lambda_{cr} 
\end{cases}.
\end{equation}
where $\lambda_{cr}$ can be considered as the critical stretch marking the onset of the damage, $0 \leq c \leq 1$ regulates the maximum damage and must be chosen to allow the material to completely degrade ($d = 1$), and $\gamma \geq 0$ defines the steepness of the transition from no damage ($d = 0$) to full damage in the damage function. We note that a more general form could be defined for the damage conjugate force as $f_d(d, \Bar{\lambda}, \dot{\Bar{\lambda}})$ for a rate-dependent case. The damage conjugate force is also sensitive to $\mathcal{H}$, a history function assuming the maximum value that the argument (in this case $\Bar{\lambda}$) has obtained throughout the loading, defined as:
\begin{equation}\label{history}
\mathcal{H}(\Bar{\lambda};\mathbf{X},t)=\mathrm{max}_{s\in[0,t]}\Bar{\lambda}(\mathbf{X},s)\,.
\end{equation}
The balance law from Eq. \eqref{damage_governing1} allows rearranging Eq. \eqref{damage_conjugateforce} to obtain what is traditionally posed as a damage function in the context of GED formulations, as follows:
\begin{equation}\label{damage_function}
d = 
\begin{cases} 
0 & \text{if } \Bar{\lambda} < \lambda_{cr} \\
1 - \frac{\lambda_{cr}-1}{\mathcal{H}(\Bar{\lambda})-1}\left(1 - c + c e^{-\gamma(\mathcal{H}(\Bar{\lambda}) - \lambda_{cr})}\right) & \text{if } \Bar{\lambda} \geq \lambda_{cr} 
\end{cases},
\end{equation}
This is the solution of the strong form of the corresponding balance law from Eq. \eqref{damage_governing1}, and it also directly enforces the non-decreasing requirement for the damage state variable directly. The resulting damage function is widely used \cite{sarkar2019comparative} but was introduced here through the definition of a constitutive law for the damage conjugate force.

Several damage functions have been proposed in the literature \cite{peerlings1996gradient, verhoosel2011isogeometric}.
One could also motivate a damage function directly from statistical mechanics arguments following Mulderrig {\textit{et al.}} (2023) \citep{mulderrig2023statistical}, which will be a future task for the authors (along with utilizing a statistical mechanics-based Helmholtz free energy function $\psi(\mathbf{F})$).

\subsection{Strong and weak forms} \label{section:strong_weak_forms}
The micro-force equilibrium equation and its associated traction boundary condition in Eq. \eqref{chain_governing} can be re-written as
\begin{equation} \label{chain_strong_boundary_1}
\begin{split}
    h_{nl}\left[\lambda_{ch} - \Bar{\lambda} + \nabla\cdot\left(g(d)\ell^2 \nabla\Bar{\lambda}\right)\right] = 0\quad \text{in} \quad \Omega_0, \\
    h_{nl}g(d)\ell^2\nabla\Bar{\lambda}\cdot\bm{n}_0 = \hat{\iota} \quad \text{on} \quad \partial_{N}{\Omega}_0.
\end{split}
\end{equation}
Considering that $h_{nl}$ is a positive value, we can therefore divide out $h_{nl}$ in both Eqs. $\eqref{chain_strong_boundary_1}_1$ and $\eqref{chain_strong_boundary_1}_2$ \citep{peerlings2004thermodynamically}. Also, according to Sarkar \textit{et al.} (2020) \cite{sarkar2020thermo}, and consistent with our interpretation of the physical process, $\hat{\iota}$ is negligible. These simplifications result in the following equations,
\begin{equation}
\begin{split}
    \lambda_{ch} - \Bar{\lambda} + \nabla\cdot\left(g(d)\ell^2 \nabla\Bar{\lambda}\right) = 0\quad \text{in} \quad \Omega_0, \\
    \nabla\Bar{\lambda}\cdot\bm{n}_0 = 0 \quad \text{on} \quad \partial_{N}{\Omega}_0.
\end{split}
\end{equation}

By incorporating the constitutive equations into the governing equations presented in Eq. \eqref{mechanical_equilibrium_governing} and Eq. \eqref{chain_governing}, we arrive at the following strong forms:
\begin{equation}
\begin{split}\label{strong_forms}
    \nabla\cdot\left[a(d)\frac{\partial\psi}{\partial\mathbf{F}} - b(d) p J \mathbf{F}^{-T}\right] = \mathbf{0} \quad \text{in} \quad \Omega_0, \\
    - b(d)(J-1) -\frac{p}{\kappa} = 0 \quad \text{in} \quad {\Omega}_0, \\
    \Bar{\lambda} - \lambda_{ch} - \nabla\cdot\left(g(d)\ell^2 \nabla\Bar{\lambda}\right) = 0\quad \text{in} \quad \Omega_0,
\end{split}
\end{equation}
with the corresponding boundary conditions
\begin{equation}\label{BC}
\begin{split}
\bm{u} = \Bar{\bm{u}} \quad \text{on} \quad \partial_{D}{\Omega}_0, \\
\mathbf{P}\cdot\bm{n}_0 = \bm{T} \quad \text{on} \quad \partial_N\Omega_0, \\
\Bar{\lambda}= \Bar{\Bar{\lambda}} \quad \text{on} \quad \partial_{D}{\Omega}_0, \\
\nabla\Bar{\lambda}\cdot\bm{n}_0 = 0 \quad \text{on} \quad \partial_{N}{\Omega}_0.
\end{split}
\end{equation}

As mentioned earlier, the traditional strain-based GED and our previous stretch-based GED \citep{mousavi2024evaluating} suffer from the non-physical broadening of the damage zone. Even though the relaxation function $g(d)$ has been introduced in Eq. \eqref{g} to counteract this issue, previous works in the literature (e.g. \citep{wosatko2021comparison,wosatko2022survey}) show no concrete evidence that this is necessarily sufficient. To alleviate this issue, one has to note that the forcing term in Eq. $\eqref{strong_forms}_3$ is the local chain stretch $\lambda_{ch}$. The physical meaning of $\lambda_{ch}$ has to be considered in the context of polymer chain statistical mechanics. As damage evolves in the damage region, $\lambda_{ch}$ will keep increasing as a crack forms and continues to open (consistent with monotonic loading). This indicates that even when a fracture surface has formed, the nonlocal driving force of Eq. $\eqref{strong_forms}_3$ keeps increasing, leading to a non-physical evolution of the nonlocal chain stretch $\Bar{\lambda}$ consistent with the observation of damage zone broadening. Polymer chain statistical mechanics provide insight into the chain scission process. As extensively discussed in our previous work \citep{mulderrig2023statistical} along with the work of Wang \textit{et al.} (2019) \cite{wang2019quantitative} and citations therein, chain scission is treated as a thermally activated process, where the energy barrier for chain scission depends on the applied chain stretch. Consequently, there exists a maximal chain stretch value where the activation energy barrier for scission goes to zero, beyond which no chains can survive. In this work, we denote this maximal chain stretch value as $\lambda_{ch}^{\text{max}}$. We postulate here that $\lambda_{ch}^{\text{max}}$ is the physically meaningful upper bound for $\lambda_{ch}$ that can influence the evolution of $\Bar{\lambda}$.\footnote{Note that $\lambda_{ch}^{\text{max}}$ is distinct from $\lambda_{cr}$, where the former is an upper bound to the allowable chain stretch that directly informs the network-level nonlocal chain stretch state, and the latter indicates damage initiation.}\footnote{In other material systems, it is not as straightforward to postulate the existence of such an upper bound, but this can still indicate the importance of such a development. Moving from scalar nonlocal variables to tensorial nonlocal variables, such an upper bound can be defined from a strain/stretch-based yield/damage function.  Even though the scope of this work is not to further build on the explicit statistical mechanics framework that we have previously developed, we use this information to limit the evolution of $\Bar{\lambda}$ in the attempt to alleviate the damage zone broadening issue and thereby improve the performance of the GED framework.}

To integrate this physical modeling feature into the framework, we substitute $\lambda_{ch}$ in Eq. \eqref{psi_nloc_micmac} with $\lambda_{ch}^{\text{max}} - \ll\lambda_{ch}^{\text{max}}-\lambda_{ch}\gg$. This modification leads to an adjusted $f_{\Bar{\lambda}}$ in Eq. $\eqref{constitutives}_4$, allowing us to reformulate Eq. $\eqref{strong_forms}_3$ as follows,
\begin{equation}\label{GED_strong2}
\Bar{\lambda} - \lambda_{ch}^{\text{max}} + \ll\lambda_{ch}^{\text{max}}-\lambda_{ch}\gg - \nabla\cdot\left(g(d)\ell^2 \nabla\Bar{\lambda}\right) = 0\text{~in~}{\Omega}_0.
\end{equation}
Note the use of the Macaulay brackets, whose operation is defined as
\begin{equation}
\ll x\gg~= \begin{cases}
x,& \text{if~}x > 0 \\
0,& \text{if~}x \leq 0 \\
\end{cases}. \label{macaulay_brackets}
\end{equation}
This adjustment introduces an upper limit for $\lambda_{ch}$ within Eq. \eqref{GED_strong2}, preventing it from non-physically increasing above $\lambda_{ch}^{\text{max}}$. 

To derive the weak form, we introduce the trial functions $(\bm{u}, p, \lambda) \in (\mathbb{U}, \mathbb{P}, \mathbb{L})$, where the function spaces $\mathbb{U}$, $\mathbb{P}$, and $\mathbb{L}$ are defined as
\begin{equation}\label{trial_spaces}
\begin{split}
\mathbb{U} = \{\bm{u} \in H^1(\Omega_0); \bm{u} = \Bar{\bm{u}} \text{ on } \partial_D\Omega_0\}, \\
\mathbb{P} = \{p \in L^2(\Omega_0)\}, \\
\mathbb{L} = \{\lambda \in H^1(\Omega_0); \lambda = \Bar{\Bar{\lambda}} \text{ on } \partial_D\Omega_0\},
\end{split}
\end{equation}
where $H^1$ is the first order Sobolev space, and $L^2$ is the second order Lebesgue space. Furthermore, we define the test functions $(\bm{v}, q, \beta) \in (\mathbb{V}, \mathbb{Q}, \mathbb{B})$, where the function spaces $\mathbb{V}$, $\mathbb{Q}$, and $\mathbb{B}$ are defined as
\begin{equation}\label{test_spaces}
\begin{split}
\mathbb{V} = \{\bm{v} \in H^1_0(\Omega_0); \bm{v} = \bm{0} \text{ on } \partial_D\Omega_0\}, \\
\mathbb{Q} = \{q \in L^2(\Omega_0)\}, \\
\mathbb{B} = \{\beta \in H^1_0(\Omega_0); \beta = 0 \text{ on } \partial_D\Omega_0\}.
\end{split}
\end{equation}

Following standard procedures, we obtain the weak form as
\begin{equation}\label{weak_form_GED}
\begin{split}
    \int_{\Omega_0} \left[a(d)\frac{\partial\psi}{\partial\mathbf{F}} - b(d) p J \mathbf{F}^{-T}\right]  : \nabla\bm{v} \,dV -\int_{\partial_N\Omega_0}{\Bar{\bm{T}}}\cdot\bm{v} \,dA & = 0, \\
    -\int_{\Omega_0}  \left(b(d)\left(J-1\right) + \frac{p}{\kappa}\right)q \,dV & = 0, \\ 
    \int_{\Omega}\Bar{\lambda}\beta \,dV - \int_{\Omega}\left[\lambda_{ch}^{\text{max}} + \ll\lambda_{ch}^{\text{max}}-\lambda_{ch}\gg\right] \beta\,dV + \int_{\Omega}g(d)\ell^2\nabla\Bar{\lambda}\cdot\nabla\beta \,dV & = 0.
\end{split}
\end{equation}

\section{Numerical implementation}\label{Section:NumericalImplementation}
Our numerical implementation of the nonlocal chain stretch-based GED model is built upon the \texttt{FEniCS} finite element framework \cite{alnaes2015fenics}, utilizing the automatic differentiation capabilities provided by the Unified Form Language (UFL). The associated Python scripts are available on GitHub\footnote{\url{https://github.com/MMousavi98/Improved_GED}}. Due to the near-incompressibility of the problem, a mixed finite element formulation is employed. Following the approach of \cite{li2020variational}, the displacement, pressure, and nonlocal chain stretch fields are discretized as
\begin{equation}\label{eq_displacement}
\bm{u}(\bm{x})= \sum_{i=1}^{N} N_i(\bm{x}) \bm{u}_i, \quad\quad\quad\quad
p(\bm{x})= \sum_{i=1}^{N} N_i(\bm{x}) p_i, \quad\quad\quad\quad
\Bar{\lambda}(\bm{x})= \sum_{i=1}^{N} N_i(\bm{x}) \Bar{\lambda}_i.
\end{equation}
Here, \(N_i(\bm{x})\) denote the shape functions associated with node \(i\), while \(\bm{u}_i\), \(p_i\), and \(\Bar{\lambda}_i\) are the corresponding nodal values for the displacement, pressure, and nonlocal chain stretch fields, respectively. To address potential oscillations due to the inf-sup condition in the coupled Eqs. $\eqref{weak_form_GED}_1$ and $\eqref{weak_form_GED}_2$, a Taylor-Hood finite element space is adopted, utilizing quadratic interpolation for the displacement field and linear interpolation for the pressure and nonlocal chain stretch fields.

Assuming quasi-static loading conditions, we implement a staggered solution scheme that iterates over a load-ramping function. The scheme consists of an outer loop that monitors convergence between successive iterations and two inner loops for solving the staggered problem. In the first inner loop, the mechanical equilibrium and Lagrange multiplier equations (i.e. Eqs. $\eqref{weak_form_GED}_1$ and $\eqref{weak_form_GED}_2$) are solved simultaneously using the \texttt{FEniCS} built-in nonlinear solver \texttt{SNES} (with the \texttt{newtontr} method\footnote{\texttt{newtontr} solves nonlinear systems iteratively using a truncated Newton-Raphson method. At each step, the linearized system is solved approximately with an iterative solver, truncating the process based on a specified tolerance for the residual reduction or a maximum number of iterations.}). This is completed while holding $\Bar{\lambda}$ fixed, denoted as $\Bar{\lambda}_{j-1}$, where $j$ is the iteration counter for the outer loop. The \texttt{SNES} solver employs a critical point \texttt{cp} line search, with absolute, relative, and solution tolerances set to $10^{-10}$, and a maximum of 300 iterations. In the second inner loop, we solve the nonlocal chain stretch equation (Eq. $\eqref{weak_form_GED}_3$) using the nonlinear constrained solver (\texttt{SNES} with the \texttt{vinewtonssls} method), applying the same tolerances and iteration limits while keeping the displacement and pressure fields fixed at $\mathbf{u}_j$ and $p_j$. The outer loop checks for convergence by evaluating whether the supremum norm error $|\Bar{\lambda}_j - \Bar{\lambda}_{j-1}|_{\infty}$ is below $2\times10^{-3}$, serving as the third convergence criterion. If this condition is met, the simulation advances to the next loading step, with $\mathbf{u}$, $p$, and $\Bar{\lambda}$ used as the initial values for the subsequent iteration. The maximum number of iterations for this criterion is also set to 300.

The energy release rate of the material is calculated through a domain J-integral \citep{li1985comparison}. This approach has previously been developed and utilized by the authors in a multiphysics setting for large deformation poroelasticity in elastomers \cite{bouklas2015effect}. This method was also utilized more recently for validation of fracture energy predictions in phase-field modeling \citep{mousavi2024evaluating}. Here, we note that as we are working with a damage problem, similar to small-scale yielding, the J-integral will be path-independent, as long as the damaged domain in the polymer does not intersect with the curve (or domain) on which the J-integral is computed on. 

\section{Results and Discussion}\label{Section:results}
In the following simulations, the normalized first P-K stress and hydrostatic pressure are distinguished with an overtop hat $\widehat{\left(\bullet\right)}$ and expressed as
\begin{equation}
    \mathbf{\widehat{P}} = \frac{\mathbf{P}}{\mu}, \quad \widehat{p} = \frac{p}{\mu}.
\end{equation}
All lengths and displacements are normalized with respect to the characteristic dimension of the domain in each problem.

\begin{figure}[h!]
    \centering
    \includegraphics[width=0.8\linewidth]{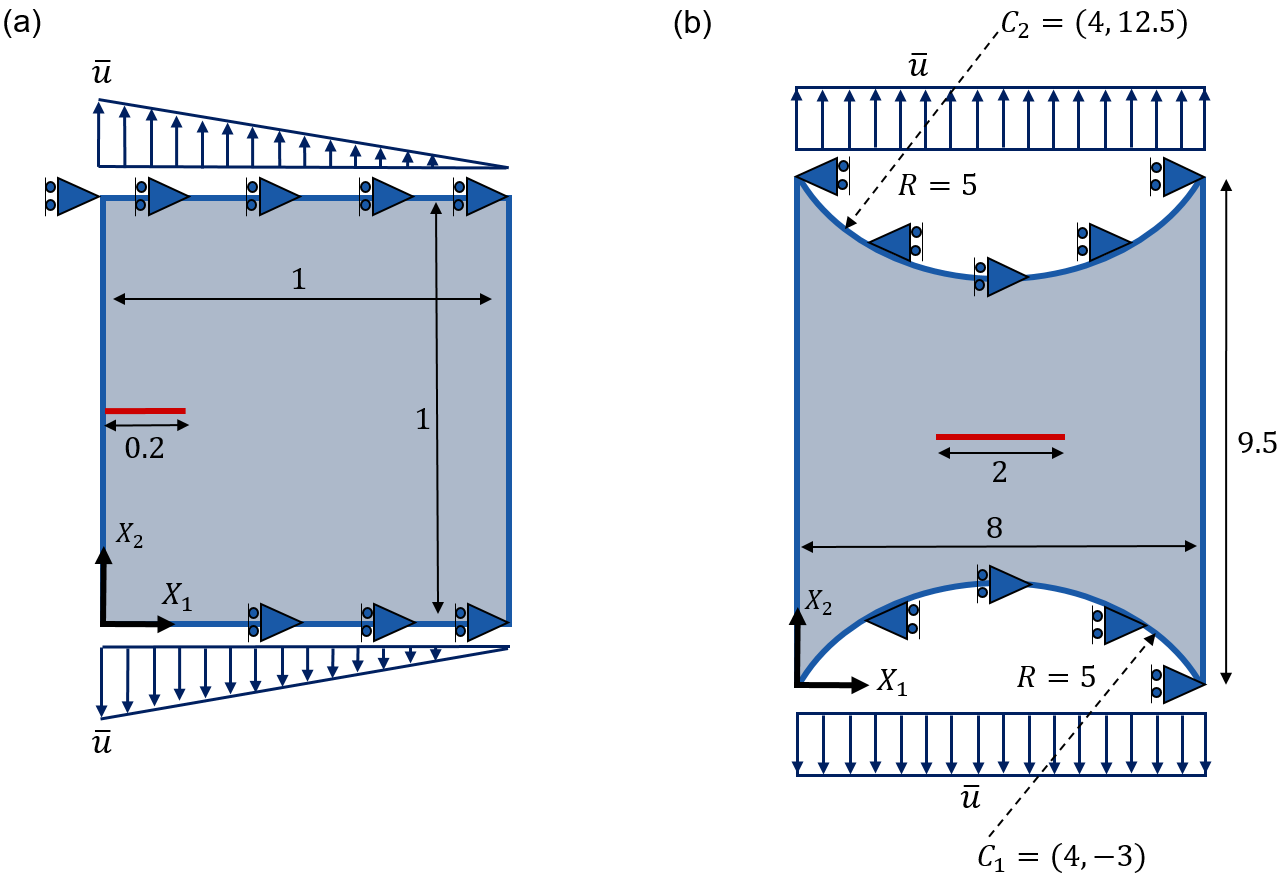}
    \caption{a) Setup of the first boundary value problem, and b) setup of the second boundary value problem.}
    \label{fig:loading}
\end{figure}

For this study, we model two different boundary value problems. In the first, we analyze a $1\times 1$ square domain (the characteristic dimension for normalization is the length of the side) in plane strain conditions, subjected to a triangular-type displacement, as illustrated in Fig. \ref{fig:loading} (a). Both the top and bottom surfaces are held fixed in the $X_1$ direction. However, on the top surface, the $X_2$ displacement component is defined as $\Bar{u}_2=0.35(1-X_1)$ (where the origin is located at the lower left corner of the domain). Conversely, the $X_2$ displacement component on the bottom surface is $\Bar{u}_2=-0.35(1-X_1)$. The traction forces are set to zero on the left and right surfaces of the domain.

In the second boundary value problem, we model crack propagation in an $8\times 9.5$ trapezoid-like domain in plane strain conditions, subjected to a uniform displacement on the top and bottom surfaces, as depicted in Fig. \ref{fig:loading} (b). The undeformed length of the crack is 2 (the characteristic dimension for normalization is the half-length of the crack). The center of the circular arcs defining the curved top and bottom surfaces are each located at $(4, 12.5)$ and $(4, -3)$, respectively, with a radius $R=5$. Similar to the first boundary value problem, the top and bottom surfaces are each held fixed in the $X_1$ direction, and there are no traction forces on the left and right surfaces. An $X_2$ displacement component of $\Bar{u}_2=0.6$ and $\Bar{u}_2=-0.6$ is uniformly applied to the top and bottom surfaces, respectively.

\begin{figure}[h!]
    \centering
    \includegraphics[width=0.6\linewidth]{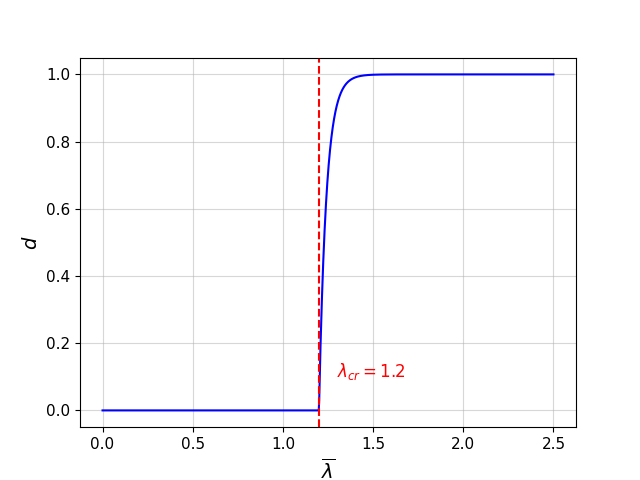}
    \caption{Damage function for $c=1$ and $\gamma=20$.}
    \label{fig:damage_function}
\end{figure}

In the simulations, we set $\ell=0.04$, $\mu=1$, and $\kappa=1000$ to maintain a ratio of $\kappa/\mu = 1 \times 10^3$, which corresponds to a Poisson ratio of $\nu = 0.499$ to ensure near-incompressibility. Additionally, in all simulations, we particularize the damage function with $c = 1$, $\gamma = 20$, and critical chain stretch $\lambda_{cr} = 1.2$, which marks the onset of network degradation. Fig. \ref{fig:damage_function} illustrates the corresponding damage function. Note that in the first boundary value problem, we have a pre-existing diffuse fracture in the domain.  This crack can be placed by setting $\Bar{\lambda}=1.5$, which results in $d=1$ according to Fig. \ref{fig:damage_function}. This initialized $\Bar{\lambda}$ field serves as the lower bound of the bounded \texttt{SNES} nonlinear solver (method: \texttt{vinewtonssls}) when we solve the nonlocal chain stretch equation (Eq. $\eqref{weak_form_GED}_3$). In the second boundary value problem, we explicitly place a narrow hole in the center of the domain with a length of 2, the thickness of $0.4$, and rounded tips with the radius of $0.2$. This problem exemplifies a case of discrete fracture, which allows the $\Bar{\lambda}$ field to be initialized to 1. The initialized $\Bar{\lambda}$ field serves as the lower bound of our bounded \texttt{SNES} nonlinear solver (method: \texttt{vinewtonssls}) when solving the nonlocal chain stretch equation.

This paper provides five numerical examples. The first four examples focus exclusively on the first boundary value problem. In the first of these examples, the damage zone broadening issue commonly observed in strain- and stretch-based GED models is replicated. The next two examples investigate two main approaches intended to mitigate damage zone broadening. We first begin by enforcing an upper bound for the nonlocal driving force, $\lambda_{ch}^{\text{max}}$, in Eq. \eqref{GED_strong2} (here, we neglect the effect of relaxation function in the equation). Then, in the following example, we activate the effect of the relaxation function $g(d)$ in Eq. \eqref{GED_strong2} and thoroughly investigate its effect on the simulation. We consider these two approaches as modeling features to Eq. \eqref{GED_strong2} intended to alleviate the broadening issue. The fourth overall example compares the results of our proposed model with those obtained from the phase-field method that has been previously validated with macroscopic experimental data. Lastly, the final example is based on the second boundary value problem and aims to demonstrate the capabilities of our model in a different setting.

\subsection{Describing the damage zone broadening issue} \label{example1}

\begin{figure}[h!]
    \centering
    \includegraphics[width=0.7\linewidth]{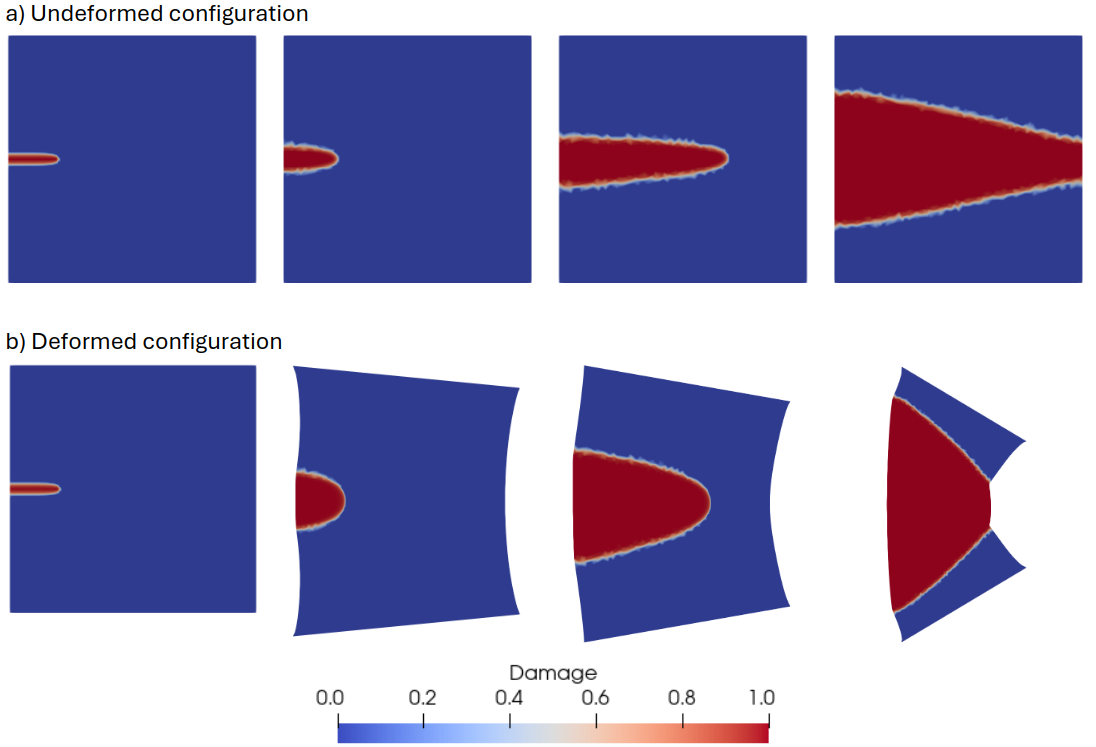}
    \caption{The result of the GED simulation for the first boundary value problem in the reference (top row) and current (bottom row) configurations at steps $1$, $17$, $30$, and $99$ (from left to right, respectively). In this particular simulation, the effect of the relaxation function $g(d)$ is neglected (i.e., $g(d)=1$) and the nonlocal driving force is unbounded.}
    \label{fig:broadening_contours}
\end{figure}

We begin by describing the issue of damage zone broadening widely observed in strain- and stretch-based GED models, as illustrated in Fig. \ref{fig:broadening_contours}. In this case, we neglect the effect of the relaxation function from Eq. \eqref{g} and fix it to $g(d)=1$. Additionally, Eq. $\eqref{strong_forms}_3$ is utilized instead of Eq. \eqref{GED_strong2}, allowing $\lambda_{ch}$ to increase the nonlocal driving force for Eq. $\eqref{strong_forms}_3$ in an unbounded manner. Note that in all contour plots in this paper, the top row represents the reference configuration, while the bottom row shows the current configuration; this permits a clear observation of damage zone broadening in the reference configuration. Clearly, the damage zone in Fig. \ref{fig:broadening_contours} is observed expanding both along the direction of crack propagation and perpendicular to it. Physically speaking, damage zone broadening is expected to take place near the vicinity of the crack tip in a process zone defined by the characteristic nonlocal length scale $l$. Such damage zone broadening is not expected to occur in the wake of the crack tip. However, the results indicate that the numerical implementation leads to extreme damage zone broadening ahead and especially behind the crack tip during crack propagation. 

This discrepancy arises because the nonlocal driving force in Eq. $\eqref{strong_forms}_3$ depends on $I_1$ (as per Eqs. $\eqref{invariants}_1$ and \eqref{Lambda_chain_ch}), which stems from the macroscopic deformation. The diffuse representation of the crack discontinuity leads to exceedingly high $I_1$ values in that damage region, which further leads to a non-physical increase in $\lambda_{ch}$. According to Eq. $\eqref{strong_forms}_3$, an increase in the nonlocal driving force (from increasing $\lambda_{ch}$) leads to a corresponding rise in the nonlocal chain stretch $\Bar{\lambda}$, which in turn causes damage $d$ to increase as well, following Eq. \eqref{damage_function}. This snowball effect ultimately leads to damage zone broadening. 

\newpage
\subsection{Employing an upper bound for the nonlocal driving force}\label{example2}

\begin{figure}[h!]
    \centering
    \includegraphics[width=0.6\linewidth]{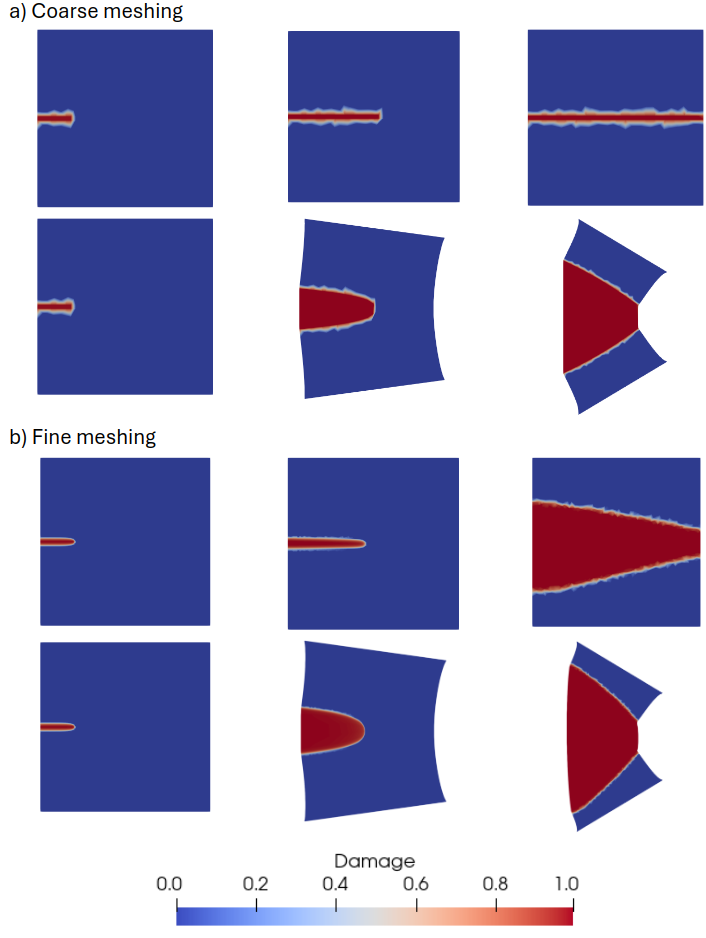}
    \caption{The result of the GED simulation for the first boundary problem at steps $1$, $24$, and $99$ (from left to right, respectively) using a) coarse meshing and b) fine meshing. The meshings are shown in Fig. \ref{fig:First_constraint_meshing}. In this particular simulation, the effect of the relaxation function $g(d)$ is neglected (i.e., $g(d)=1$), but the nonlocal driving force is bounded.}
    \label{fig:First_constraint_contours}
\end{figure}

\begin{figure}[h!]
    \centering
    \includegraphics[width=0.7\linewidth]{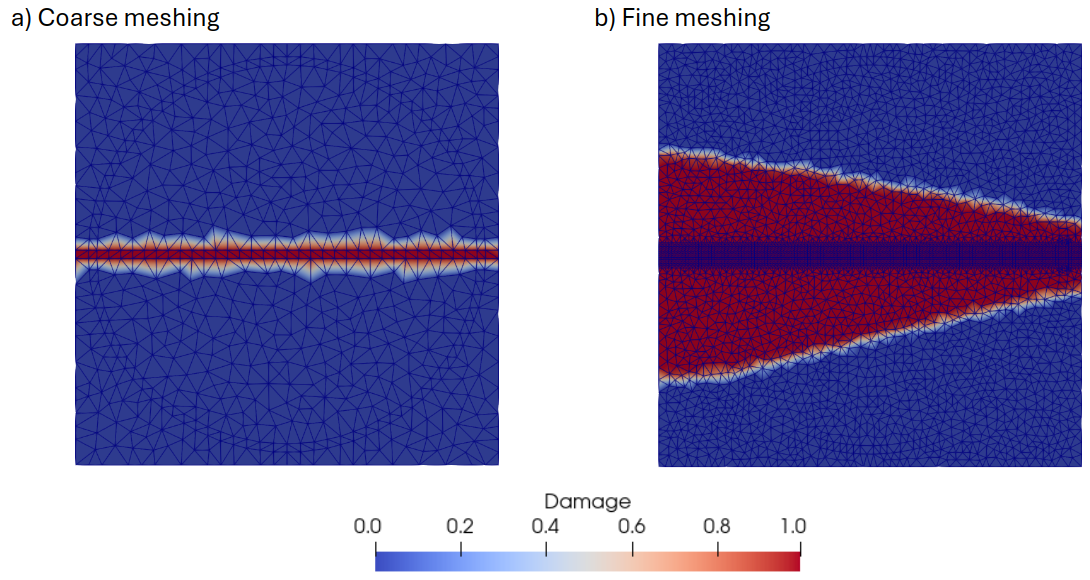}
    \caption{Depictions of the a) coarse mesh and b) fine mesh used in the GED simulation for the first boundary value problem (with full results displayed in Fig. \ref{fig:First_constraint_contours}).}
    \label{fig:First_constraint_meshing}
\end{figure}

To prevent the broadening of the damage zone, the general range of $\lambda_{ch}$ that influences damage and ultimately fracture has to be consistent with the principles of polymer chain statistical mechanics and mechanochemical chain scission \citep{mulderrig2023statistical}. This is the departure point for introducing an ostensible maximum chain stretch, $\lambda_{ch}^{\text{max}}$, intended to bind the nonlocal driving force in Eq. $\eqref{strong_forms}_3$. We constrain the nonlocal driving force with $\lambda_{ch}^{\text{max}}$ via the methodology discussed in Section \ref{section:strong_weak_forms}. At this point, the effect of the relaxation function is still neglected and remains fixed at $g(d)=1$. As demonstrated in Figs. \ref{fig:First_constraint_contours} (a) and \ref{fig:First_constraint_meshing} (a), binding the nonlocal driving force using $\lambda_{ch}^{\text{max}}=2.7$ successfully prevents damage zone broadening for the specific example. Note that this simulation was performed on a relatively coarse mesh (with the element size of $h=0.02$ to $h=0.03$ in the damage region). To evaluate the sensitivity of the model to mesh size, we repeated the same simulation on a finer mesh (with the element size of $h=0.005$), as depicted in Figs. \ref{fig:First_constraint_contours} (b) and \ref{fig:First_constraint_meshing} (b). The damage contours in Fig. \ref{fig:First_constraint_contours} (b) in particular reveal that this new modeling feature initially controls the damage zone broadening rather effectively. However, as the domain becomes further deformed, damage begins to unnaturally spread and broaden once more. This reveals that enforcing an upper bound for the nonlocal driving force is itself not enough to mitigate damage broadening. 

The issue stems from the gradient term in Eq. \eqref{GED_strong2}, which incorporates nonlocal effects into the equation. In this context, $\Bar{\lambda}$ is influenced not only by the local state of $\lambda_{ch}$ but also by the stretch state of adjacent nodes. Thus, even though constraining $\lambda_{ch}$ prevents it from exceeding $\lambda_{ch}^{\text{max}}$ (even inside the damage zone), the nonlocal influence on $\Bar{\lambda}$ still remains (through the nonlocal gradient term), and can sometimes dominate. This nonlocal gradient effect is dependent on the length scale $\ell$. In the coarse mesh, the element sizes are nearly equal to $\ell$, which inhibits the transfer of nonlocal effects between elements. In contrast, the finer mesh (with element sizes an order of magnitude less than $\ell$) allows the nonlocal gradient effects to influence surrounding elements, ultimately leading to damage broadening. As a result, the first modeling feature is indeed effective, but not sufficient. Before going to the next part to further address this issue, it is insightful to investigate the effect of varying $\lambda_{ch}^{\text{max}}$ on the model response.

\begin{figure}[h!]
    \centering
    \includegraphics[width=0.8\linewidth]{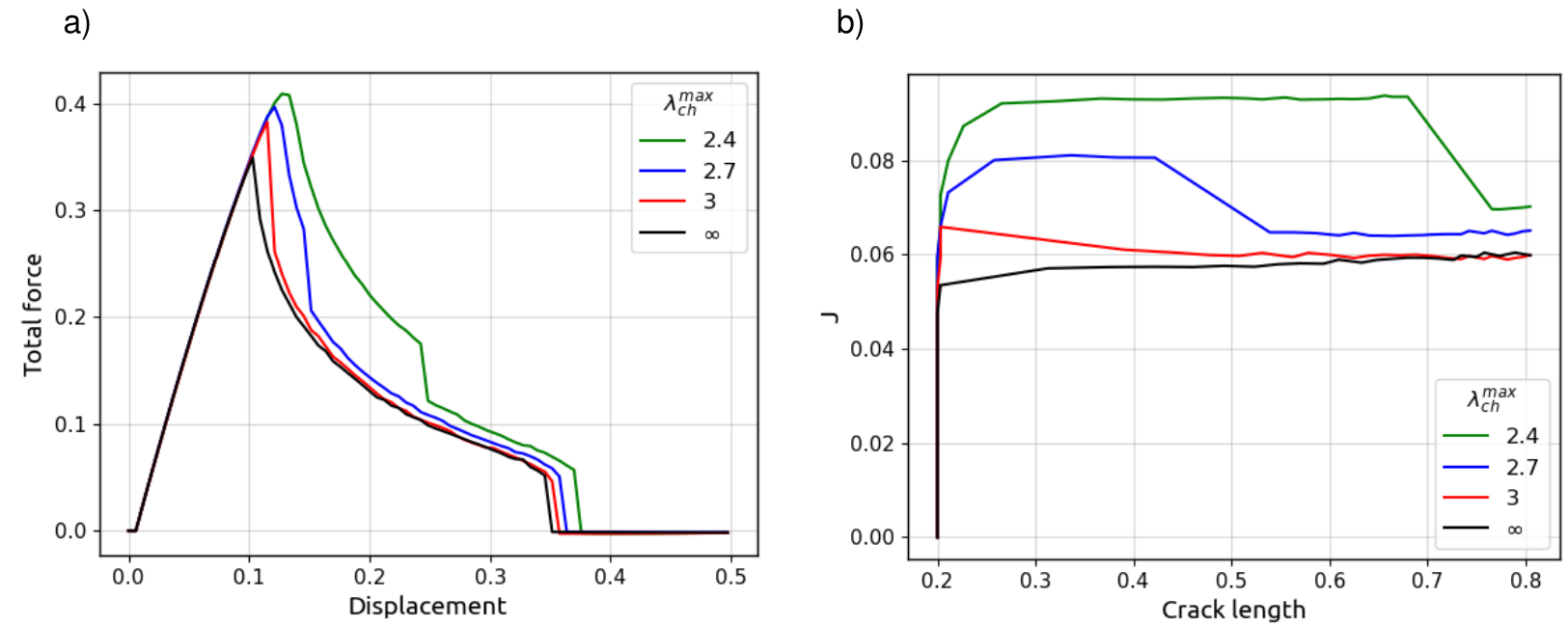}
    \caption{The effect of employing an upper bound for the nonlocal driving force, as per varying $\lambda_{ch}^{\text{max}}$, on the a) force-displacement curve and b) energy release rate as obtained from the domain J-integral.}
    \label{fig:First_constraint_graphs}
\end{figure}

By selecting a set of values for $\lambda_{ch}^{\text{max}}$, Fig. \ref{fig:First_constraint_graphs} showcases the overall impact of $\lambda_{ch}^{\text{max}}$ on the force-displacement response and fracture energy predictions as evaluated by the domain J-integral\footnote{There are limitations to the J-integral in this damage problem, following similar concepts to small scale yielding. As long as the damage zone does not intersect with the domain on which we evaluate the J-integral, then the evaluation will be path independent. Once the damage zone, or the crack itself, intersects with the J-integral evaluation domain, then the results are not meaningful. As such, henceforth in this work, there is always a point during crack propagation when the damage zone will intersect with our J-integral domain, noting the extent of the validity of the reported J-integral curve. Given that the damage zone is rather localized in this work, there is a significant extent of crack propagation in our examples henceforth prior to that point. For the cases shown in the examples below, the point where the damage zone intersects with the J-integral domain is when the crack length is approximately 0.75.}. Decreasing $\lambda_{ch}^{\text{max}}$ effectively decreases the nonlocal driving force for damage, and allows the material to support greater loads, resulting in a higher critical energy release rate. As previously mentioned, while this feature is initially effective, broadening eventually occurs, leading to a noticeable ``jump'' in both plots. The smaller the value of $\lambda_{ch}^{\text{max}}$, the more delayed this jump occurs in the deformation history. It is important to note that this broadening and the resulting jump each lack physical justification; both phenomena do not align with our expectations and previous experimental results that agree with phase-field model predictions, indicating a need for further investigation into this issue.

\subsection{Incorporating the relaxation function} \label{example3}

\begin{figure}[h!]
    \centering
    \includegraphics[width=0.7\linewidth]{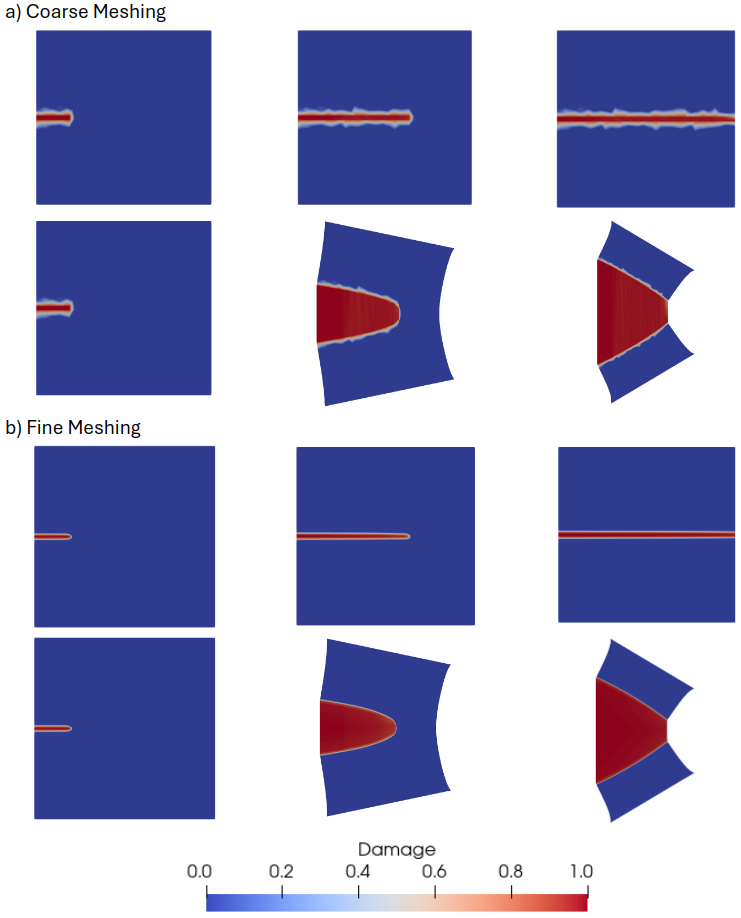}
    \caption{The result of the GED simulation for the first boundary problem at steps $1$, $24$, and $99$ (from left to right, respectively) using a) coarse meshing and b) fine meshing. In this particular simulation, the relaxation function $g(d)$ is activated and the nonlocal driving force is bounded.}
    \label{fig:both_constraint_eta0_contours}
\end{figure}

The aim here is to reduce the nonlocal influence from nodes within the damage zone as they approach a fully damaged state. This can be accomplished by using the relaxation function $g(d)$, which was introduced in Eq. \eqref{psi_nloc_grd}.
The relaxation function depends on the damage field and consequently diminishes nonlocal chain stretch interactions as a material point deteriorates through increasing damage. At the limit of full damage, all such nonlocal interactions are turned off. In this example, we set $\lambda_{ch}^{\text{max}} = 2.7$ to bind the nonlocal driving force, as in the previous example. Using the relaxation function from Eq. \ref{g},  we choose the relaxation function parameter $m = 0.28$ (this choice will be discussed later in this section). This selection ensures a sharp relaxation of the nonlocal effects upon significant damage accumulation as seen in Fig. \ref{fig:g_plot}, consistent with our physical understanding of the process. The results, as illustrated in Fig. \ref{fig:both_constraint_eta0_contours}, reveal that there is no broadening of the damage zone when using either the coarse mesh or the fine mesh. A more comprehensive investigation studying the mesh sensitivity of the simulation results with this particular model is presented in Appendix \ref{appendix1}. 

Fig. \ref{fig:profiles} presents five different 2D contour plots, and the corresponding profiles normal to the fracture surface at $X_1 = 0.3$ for $d$, $\Bar{\lambda}$, $\lambda_{ch}$, $p$, and $J=\textrm{det}(\mathbf{F})$. 
From the normal profiles of the quantities of interest we can understand the model behavior in the crack vicinity. The damage field is rather confined normal to the crack direction due to the selection of the specific damage function with damage initiation chain stretch $\lambda_{cr}$ (as per Eq. \eqref{damage_function}), in contrast to the nonlocal chain stretch $\Bar{\lambda}$ that decays more smoothly. The local chain stretch $\lambda_{ch}$ is significantly more localized, and follows the unloading of the material. Moreover, the $J$ contour reflects the volume change of the body. Due to the near-incompressibility constraint, $J$ remains equal to $1$ everywhere in the domain except in the fully damaged region, signifying that the volume is preserved outside the damaged zone. The pressure bubble and oscillation in the damage region are expected from the mixed formulation presented here, and consistent with our previous perturbed Lagrangian formulation for the phase field extensively discussed in Ang {\textit{et al.}} (2022) \citep{ang2022stabilized}.
\begin{figure}[h!]
    \centering
    \includegraphics[width=0.7\linewidth]{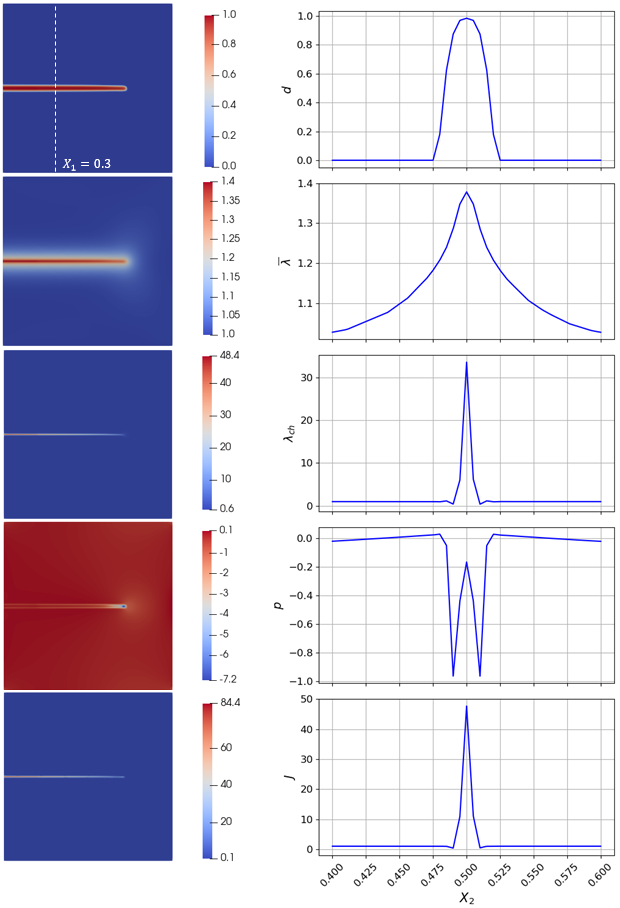}
    \caption{Damage $d$, nonlocal chain stretch $\Bar{\lambda}$, chain stretch $\lambda_{ch}$, hydrostatic pressure $p$, and the volume change $J=\textrm{det}(\mathbf{F})$ profiles for the first boundary value problem over the undeformed configuration (left column), and from top to bottom at $X_1=0.3$ and $0.4<X_2<0.6$ (right column). In the contour plots, there is a minor interpolation error near elements of high distortion at the visualization stage (gradients are saved in a discontinuous per-element space upon completion of the solution, but then interpolated for plotting).}
    \label{fig:profiles}
\end{figure}

\begin{figure}[h!]
    \centering
    \includegraphics[width=0.8\linewidth]{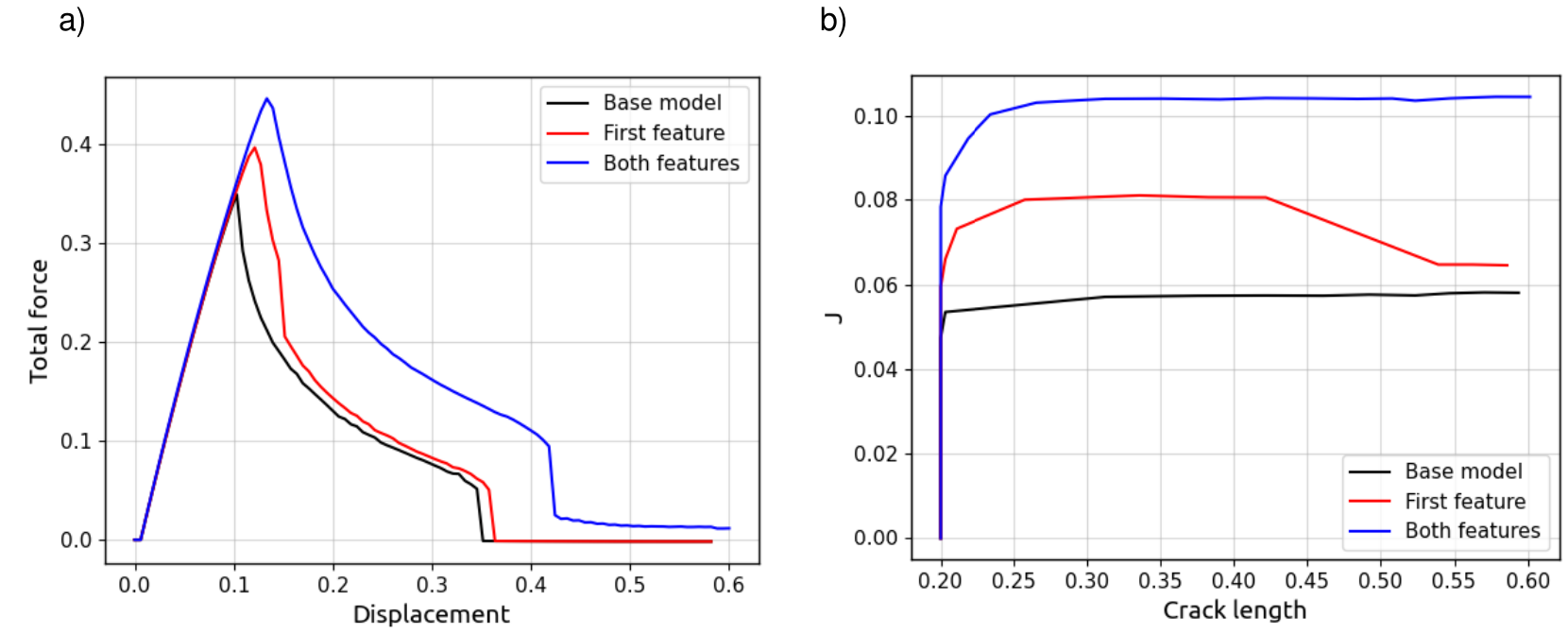}
    \caption{A comparison between the three GED modeling scenarios on the a) force-displacement curve and b) energy release rate. ``Base model'' refers to the model $g(d)=1$ and unbounded nonlocal driving force. ``First feature'' refers to the model with $g(d)=1$ and bounded nonlocal driving force. ``Both features'' refers to the model with a decaying $g(d)$ and bounded nonlocal driving force.}
    \label{fig:both_constraint_eta0_graphs}
\end{figure}

Fig. \ref{fig:both_constraint_eta0_graphs} presents a comparison of the force-displacement curves and J-integral curves for the three modeling scenarios: a simulation utilizing (i) $g(d)=1$ with an unbounded nonlocal driving force, (ii) $g(d)=1$ with a bounded nonlocal driving force (via Eq. \eqref{GED_strong2}), and (iii) an activated relaxation function $g(d)$ (via Eq. \eqref{g}) with a bounded nonlocal driving force. From this comparison, we can conclude that limiting the nonlocal driving force for damage through both modeling features ultimately results in higher toughness. Additionally, when both features are applied, the response is consistent with the form of the force-displacement response profiles often observed in elastomer fracture experiments and simulations \citep{miehe2010rate, miehe2014phase, yin2020fracture,najmeddine2024physics,ang2022stabilized}.  

\begin{figure}[h!]
    \centering
    \includegraphics[width=0.8\linewidth]{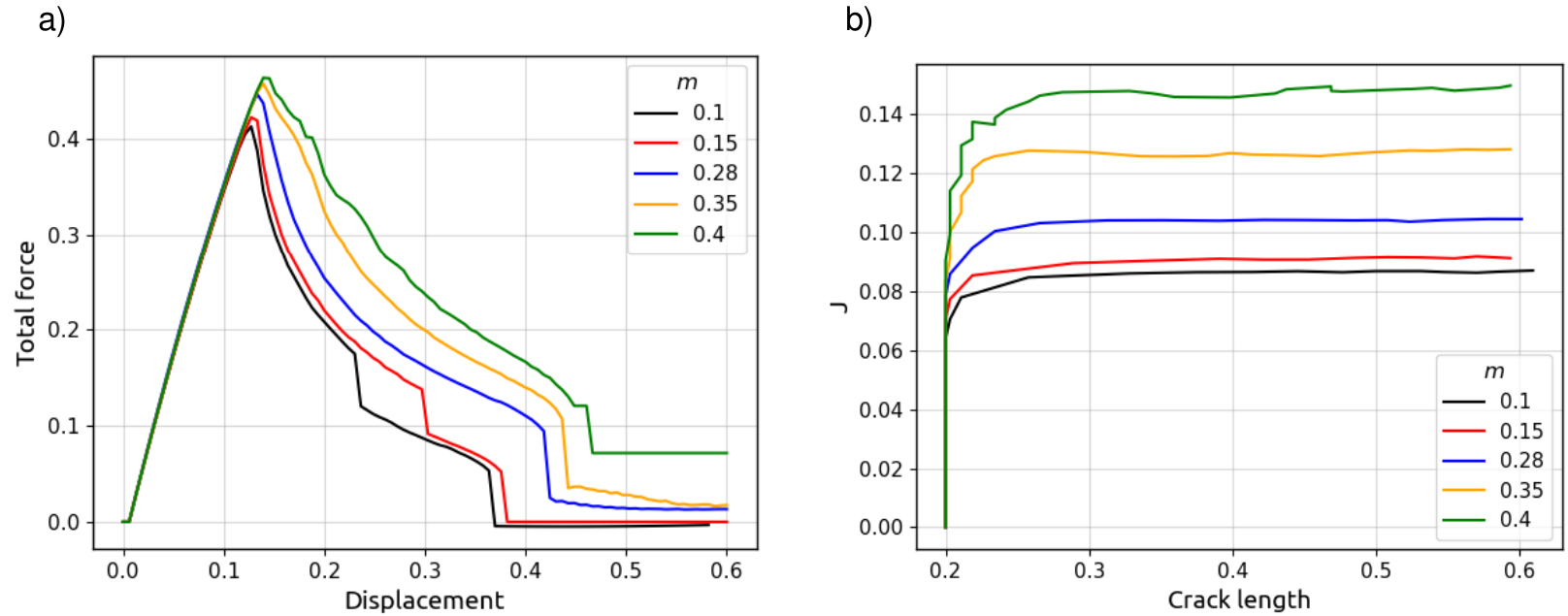}
    \caption{The effect of the relaxation function parameter $m$ on the a) force-displacement curve and b) energy release rate.}
    \label{fig:g_effect_graphs}
\end{figure}

The relaxation function parameter $m$ is crucial for the alleviation of nonlocal interactions within the highly damaged region. Nonlocal interactions are maintained in regions of little to no damage but are sharply dismantled in the vicinity of high damage. Fig. \ref{fig:g_effect_graphs} illustrates that decreasing the power of the relaxation function (i.e., decreasing $m$) reduces its effectiveness in that regard. We first focus at the response for $m=0.15$ and $m=0.1$ as suggestive of convergence at the limit $m=0$ (pointing to Fig. \ref{fig:g_plot}); we note that exploring that limit is computationally challenging. Thinking of the $m=0$ limit case, then for any damage value other than $d=1$, the nonlocal interactions are in full effect, and for cases that approach the fully damaged state, damage interpolation can trigger significant nonlocal interactions. This is showcased by a sharp drop-off in the force for both cases when the normalized displacement reaches 0.3 and 0.23, respectively (for $m=0.15$ and $m=0.1$). This drop-off is non-physical and it is also accompanied by a broadening of the damage zone. On the other hand, selecting a high value for the relaxation function parameter (e.g., $m=0.4$)  does not allow for the deterioration of the load, resulting in the development of considerable residual stresses. In light of this, $m$ should be meticulously chosen so that the relaxation function and the bounded nonlocal driving force together allow for the fracture to propagate fully while preventing the damage zone from broadening. Further investigation of this issue when using a statistical mechanics-based damage function is also necessary. All things considered, we select $m=0.28$ to diminish the broadening effect and also maintain low residual stresses in the damaged region. In the future, this is a point that can be analyzed more rigorously but is out of the scope of the current work.

\subsection{Comparison with the phase-field method}\label{example4}

\begin{figure}[h!]
    \centering
    \includegraphics[width=0.8\linewidth]{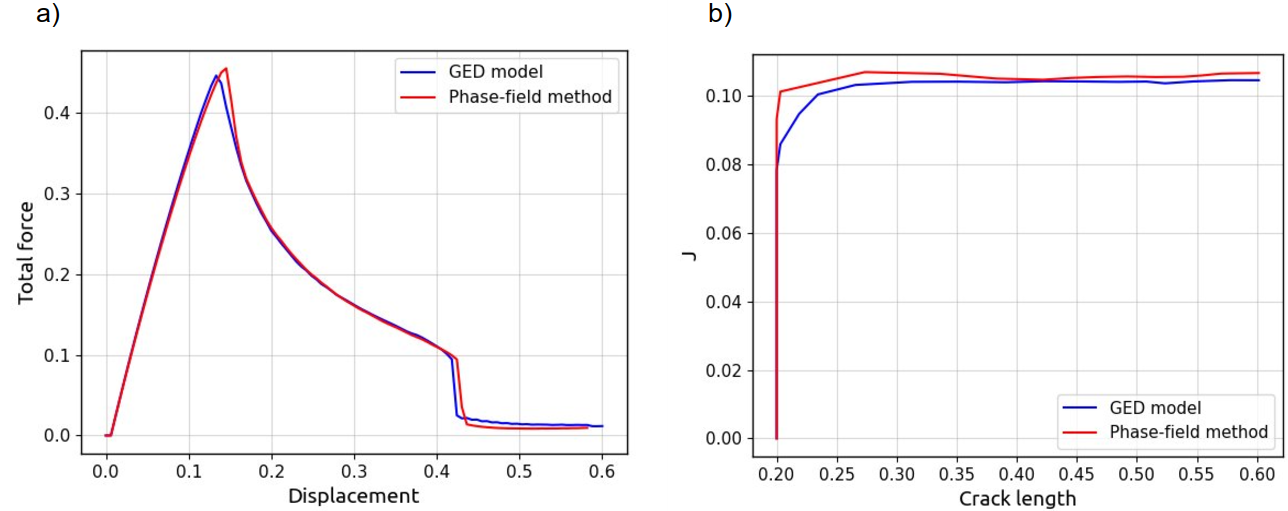}
    \caption{A comparison between the GED model and the phase-field method results based on the a) force-displacement curve and b) energy release rate.}
    \label{fig:comparison_phase_field_graphs}
\end{figure}

To further evaluate our model, we provide a comparison between the GED model and the phase-field method. The rationale behind this comparison is that the phase-field method has been carefully benchmarked on macroscopic results (e.g., crack path and load-displacement curves). However, in the process, the damage distribution (e.g., the spatial resolution of broken chains for the elastomer network) has been consistently neglected. Thus, a comparison with the phase-field method is a first step to make sure that the GED model can provide physically meaningful fracture predictions. We first perform the GED model simulation using $\lambda_{ch}^{\text{max}}=2.7$ and $m=0.28$. By using the domain J-integral approach (as described at the end of Section \ref{Section:NumericalImplementation}), we find that the critical energy release rate for the GED model simulation is $G_c=0.104$ (as seen in Fig. \ref{fig:comparison_phase_field_graphs} (b)). We then utilize this value of fracture toughness $G_c$ as the input $G_c$ value in the phase-field method simulation. Fig. \ref{fig:comparison_phase_field_graphs} compares the force-displacement curve and the energy release rate prediction (based on domain J-integral evaluation) for the GED and phase-field approaches, respectively. The mechanical response and fracture energy release response for both models comply closely with one another. Once again, the key for this comparison is that we directly utilize the energy release rate plateau for the GED model result from Fig. \ref{fig:comparison_phase_field_graphs} (b) as the $G_c$ input for the phase-field method (with no further modifications to the phase-field method). 

\begin{figure}[h!]
    \centering
    \includegraphics[width=0.55\linewidth]{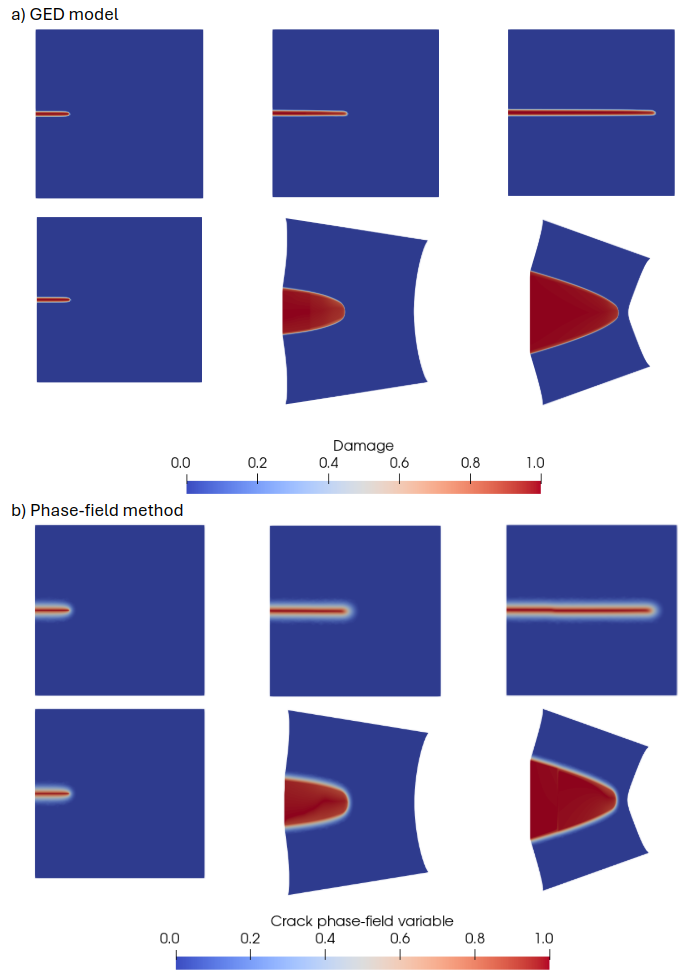}
    \caption{A comparison between a) the damage contours of the GED model and b) the crack phase-field variable contours of the phase-field method for the first boundary problem at steps $1$, $27$, and $60$ (from left to right, respectively) of the crack propagation.}
    \label{fig:comparison_phase_field_contours}
\end{figure}

Fig. \ref{fig:comparison_phase_field_contours} demonstrates the corresponding results for the damage field and the crack phase field of the GED and phase-field approaches, respectively. Notably, the contours of each field display a rather strong overlap. The importance of this striking behavior is that even though the two approaches give consistent macroscopic results during crack propagation, the GED model provides physically meaningful insight into the deterioration of the material in the vicinity of the crack relevant to the nonlocal length scale $l$. This is contrary to the phase-field method, where the deterioration is due to the numerical regularization of the sharp crack interface. The discrepancies in the response are due to the fact that specific constitutive choices were made in this work for the GED model (such as the selection of a discontinuous damage function, which is a modeling component that can be reconsidered in future studies) and the direct aim was not to perfectly match the response of the two models.


\subsection{Center-crack example}

For our final numerical vignette, we employ the second boundary value problem, as shown in Fig. \ref{fig:loading} (b). Unlike the previous cases where a diffuse fracture was introduced into the domain, a narrow hole (a rectangle with a width of $0.4$, a length of $2$, and rounded ends each with a radius of $0.2$) is here placed in the middle of the domain. From the geometrical features of the domain, we anticipate that damage and eventually cracks will emerge at both ends of the hole. Similar to the previous examples, we set $\lambda_{ch}^{\text{max}} = 2.7$ and $m = 0.28$. All of the remaining material properties are the same as described at the beginning of Section \ref{Section:results} with the exception of $\ell$. In light of the larger size of this domain (as compared to the previous examples), $\ell$ is here set to 0.06. Given the unstable nature of the loading in this particular case, we incorporate an artificial viscosity term into the formulation, assigning the artificial viscosity parameter $\eta = 0.5$ (the implications of introducing the artificial viscosity to the formulation are discussed in our previous study \citep{mousavi2024evaluating}). A detailed discussion of the artificial viscosity technique in the context of the GED formulation, along with a brief assessment of how artificial viscosity affects the simulations, can be found in Appendix \ref{appendix2}.

\begin{figure}[h!]
    \centering
    \includegraphics[width=0.7\linewidth]{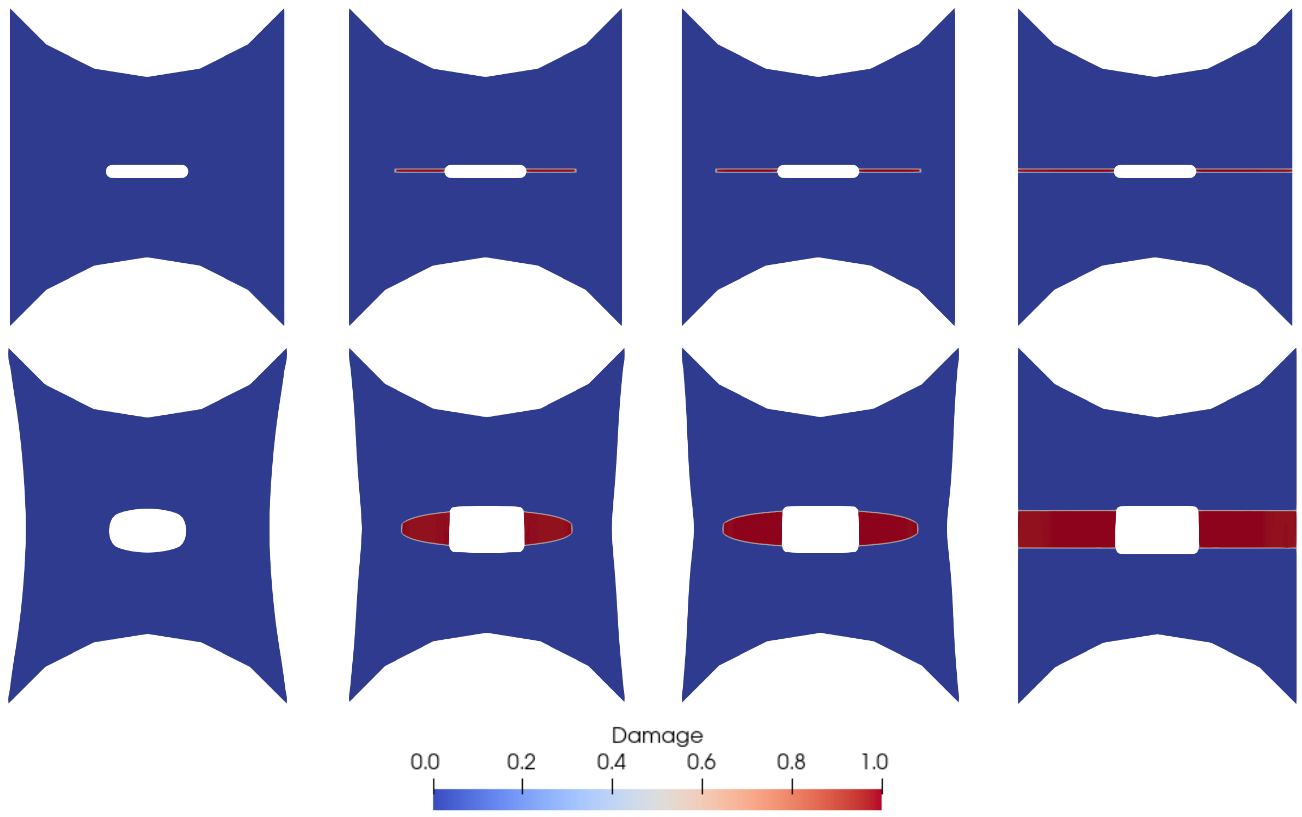}
    \caption{The damage contours from the artificial viscosity-modified GED simulation for the second boundary value problem at steps $162-165$ (from left to right, respectively). In this simulation, the artificial viscosity parameter $\eta=0.5$.}
    \label{fig:discrete_contours}
\end{figure}

\begin{figure}[h!]
    \centering
    \includegraphics[width=0.45\linewidth]{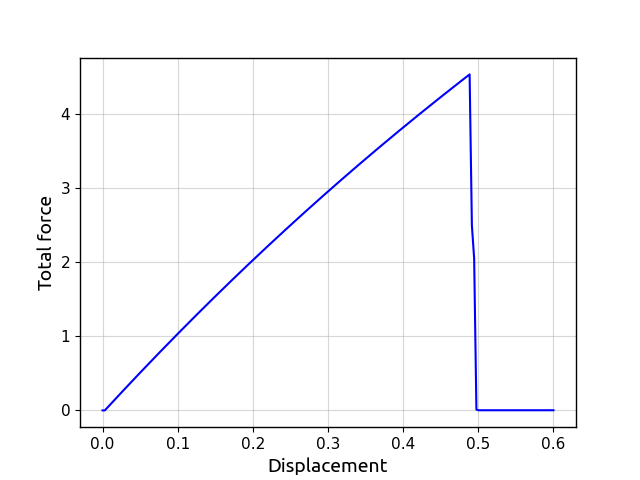}
    \caption{The force-displacement curve for the second boundary value problem.}
    \label{fig:discrete_graph}
\end{figure}

The results, displayed in Fig. \ref{fig:discrete_contours}, confirm that cracks initiate at both ends of the hole and propagate all the way to the lateral boundaries. This propagation takes place over three consecutive steps, as illustrated in the figure, emphasizing the unstable character of the problem. Additionally, the force-displacement curve in Fig. \ref{fig:discrete_graph} shows a notable contrast with the previous examples: here, the material’s load-bearing capacity decreases suddenly due to rapid crack propagation, which is unlike the previous examples where the crack grew over multiple steps.

Finally, the most critical takeaway from this example is that the proposed GED method demonstrates its capability to model an entirely different problem without encountering the damage zone broadening issue, thereby highlighting the strength and versatility of the method.

\newpage
\section{Conclusion}\label{Section:conclusion}

In this study, we developed a thermodynamically consistent stretch-based gradient-enhanced damage model to accurately simulate crack propagation and concurrent microscale damage in near-incompressible elastomers. Notably, both the physically diffuse rupture of polymer chains and the emerging formation and propagation of cracks are captured, consistent with recent experimental observations. In the literature, strain-based gradient damage models are primarily used to capture damage, and phase field models are utilized to capture fracture. In this work, we present a framework to capture the combined effects of damage and fracture in elastomers.
A key innovation of our approach is the introduction of a statistical mechanics-based upper bound of the local chain stretch value that directly influences the nonlocal chain stretch evolution. We combine this with an appropriate relaxation function that limits nonlocal effects in the fully damaged region. These two modeling features are designed to mitigate the issue of damage zone broadening which has hindered the effectiveness and reduced the accuracy of strain- and stretch-based GED formulations over the last several decades. The results were shown to remove mesh dependence and consistently retrieve the critical energy release rate, the distribution of damage, and the load-displacement response for a variety of different problems. We formulated the model by deriving the governing equations using the principle of virtual power and the constitutive laws through the Coleman-Noll procedure. The corresponding constitutive choices were made to comply with the thermodynamic constraints. Appropriate function spaces were selected for stability at the limit of incompressibility, along with the use of a perturbed Lagrangian approach for allowing volume changes in the damaged region by relaxing the incompressibility constraint. 


In future work, this model can be applied to simulate crack propagation in elastomers by utilizing constitutive laws based on polymer chain statistical mechanics. This will provide further predictive capabilities and connect to experiments, especially those where pointwise damage is visualized through light-emitting mechanophore activation during chain scission  \cite{slootman2020quantifying,slootman2022molecular, ju2024role}. These experiments allow for concurrent spatially resolved chain scission and macroscopic fracture evaluation. Enhanced understanding through theory and numerical simulations can uncover the hidden damage cascade that leads to commonly observed rate-dependent fracture phenomena in elastomers. Further extensions to the approach include the consideration of anisotropic elastic and damage responses, as well as rate-dependent damage and fracture.

\section{Acknowledgments}
SMM and NB acknowledge the support by the National Science Foundation under grant no. CMMI-2038057. JPM gratefully acknowledges the support of the National Science Foundation Graduate Research Fellowship Program under Grant No. DGE-1650441. Any opinions, findings, conclusions, or recommendations expressed in this material are those of the author(s) and do not necessarily reflect the views of the National Science Foundation. JPM also gratefully acknowledges the support of UES, Inc. (a BlueHalo Company), the Air Force Research Laboratory Materials and Manufacturing Directorate, and the National Research Council (NRC) Research Associateship Program (administered by the National Academies of Sciences, Engineering, and Medicine). The work of BT was performed under the auspices of the U.S. Department of Energy by Lawrence Livermore National Laboratory under Contract DE-AC52-07NA27344.
\section{Highlights}\label{Section:highlights}
\begin{itemize}
  \item A polymer chain stretch-based gradient-enhanced damage model is introduced to model concurrent damage and fracture in elastomers.
  \item Unlike previous stain- and stretch-based gradient-enhanced damage models, this new model does not exhibit damage zone broadening due to numerical issues.
  \item This new framework allows for fracture energy predictions, and evaluation of local degradation, starting from chain-level considerations.
  \item Degradation of the material is physically meaningful and not due to numerical regularization, as is the case of phase-field models.
\end{itemize}

\appendix
\gdef\thesection{\Alph{section}}
\makeatletter
\renewcommand\@seccntformat[1]{Appendix \csname the#1\endcsname.\hspace{0.5em}}
\makeatother
\section{Investigation of mesh size sensitivity} \label{appendix1}

\begin{figure}[h!]
    \centering
    \includegraphics[width=0.8\linewidth]{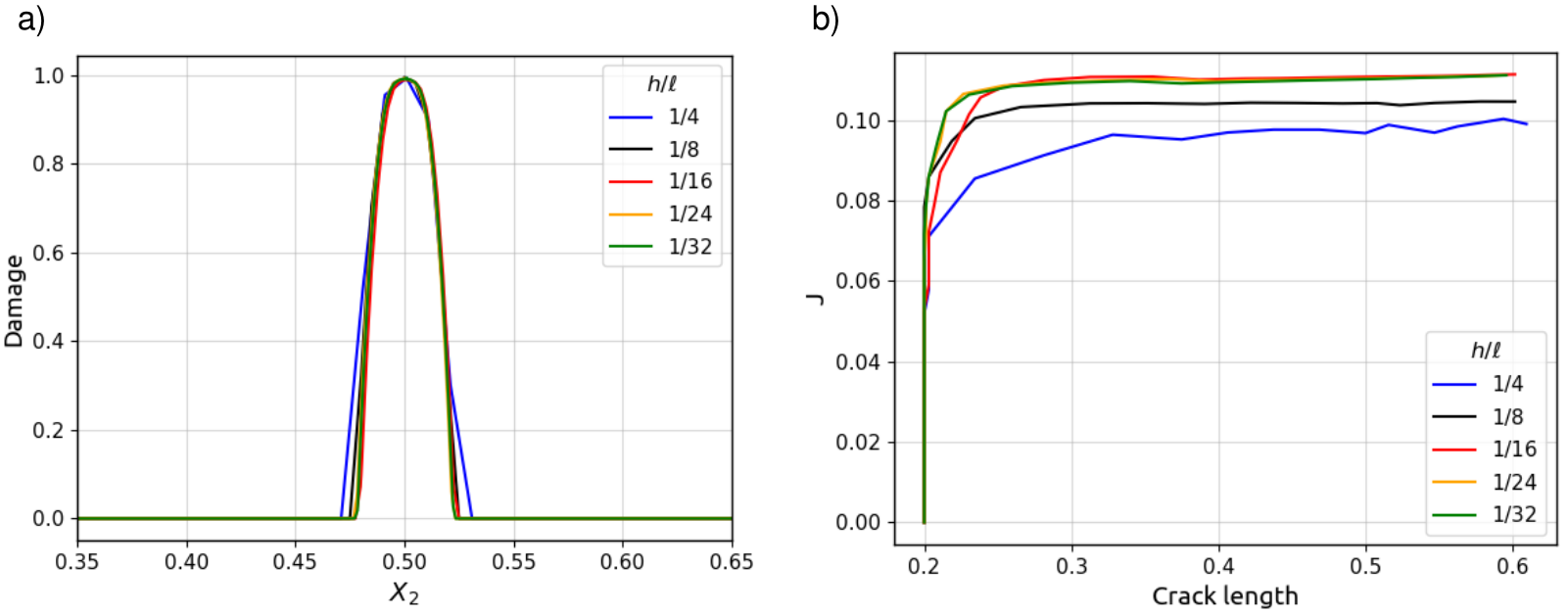}
    \caption{The effect of mesh size on the GED simulation based on a) damage profiles at $X_1=0.5$ and $0.35<X_2<0.65$ after the full propagation of the crack in the domain and b) the energy release rate profiles.}
    \label{fig:mesh_profile_J}
\end{figure}

\begin{figure}[h!]
    \centering
    \includegraphics[width=0.5\linewidth]{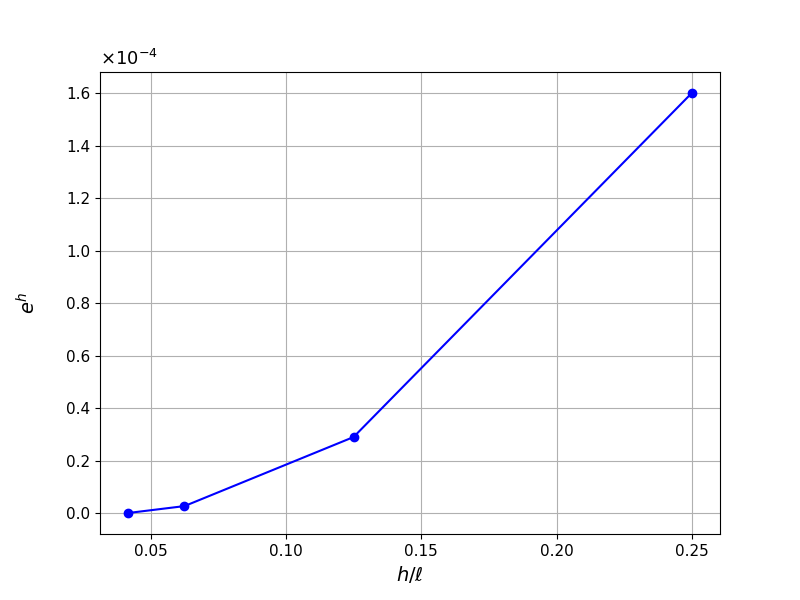}
    \caption{Convergence study of the steady state J-integral (crack length equals $0.4$) under mesh refinement.}
    \label{fig:error}
\end{figure}

In this section, we conduct a parameter study to examine the impact of mesh size on the simulation results. All model parameters remain the same as those described in Section \ref{Section:results}, with the exception of varying element sizes over five different mesh configurations. Using $\ell=0.04$ for all of the studies, the coarsest mesh has $h/\ell=1/4$ around the damage zone, and we progressively decrease this ratio to $1/8$, $1/16$, $1/24$, and $1/32$ by further refining the meshing. The results of this study are presented in Fig. \ref{fig:mesh_profile_J}. As the mesh becomes finer, both the damage profiles and the energy release rate curves show convergence. To demonstrate this further, Fig. \ref{fig:mesh_profile_J} presents the $J$ values measured when the crack has propagated to attain a crack length of $0.4$ and the J-integral shows a steady-state response (for all cases other than the coarsest mesh). Denoting the J-integral value obtained from the simulation with the finest mesh $h/\ell=1/32$ as $\hat{J}$, we compute the squared difference $e^h=(J^h-\hat{J})^2$ with respect to the remaining simulations. In Fig. \ref{fig:error} the squared error plotted against $h/\ell$ indicates proficient convergence of the fracture energy as a result of the mesh refinement.

\section{Investigating the effect of artificial viscosity} \label{appendix2}

To overcome numerical stability issues due to unstable crack growth, the use of artificial viscosity is commonly employed in phase-field formulations in order to penalize the rapid growth of the damage field \cite{miehe2010rate, ulloa2022variational, aldakheel2021multilevel, miehe2010thermodynamicPF}. We utilize the same concept of artificial viscosity here in the gradient-enhanced damage model and rewrite Eq. $\eqref{GED_strong2}$  as follows
\begin{equation}\label{GED_strong_viscosity}
\eta\dot{\Bar{\lambda}} + \Bar{\lambda} - \lambda_{ch}^{\text{max}} + \ll\lambda_{ch}^{\text{max}}-\lambda_{ch}\gg - \nabla\cdot\left(g(d)\ell^2 \nabla\Bar{\lambda}\right) = 0 \text{~in~}{\Omega}_0,
\end{equation}
along with the following boundary conditions that previously appeared in Eq. \eqref{BC},
\begin{equation}\label{BC_appendix}
\begin{split}
\Bar{\lambda}= \Bar{\Bar{\lambda}} \quad \text{on} \quad \partial_{D}{\Omega}_0, \\
\nabla\Bar{\lambda}\cdot\bm{n}_0 = 0 \quad \text{on} \quad \partial_{N}{\Omega}_0.
\end{split}
\end{equation}

By following the same process explained in Section \ref{Section:theory} for the weak forms, we can rewrite Eq. $\eqref{weak_form_GED}_3$ in light of the artificial viscosity,
\begin{equation}\label{weak_form_appendix}
\int_{\Omega}\eta\dot{\Bar{\lambda}}\beta \,dV + \int_{\Omega}\Bar{\lambda}\beta \,dV - \int_{\Omega}\left[\lambda_{ch}^{\text{max}} + \ll\lambda_{ch}^{\text{max}}-\lambda_{ch}\gg\right] \beta\,dV + \int_{\Omega}g(d)\ell^2\nabla\Bar{\lambda}\cdot\nabla\beta \,dV=0.
\end{equation}
The above equation is equivalent to Eq. $\eqref{weak_form_GED}_3$ with the exception of the first term that accounts for the artificial viscosity effect.

\begin{figure}[h!]
    \centering
    \includegraphics[width=0.6\linewidth]{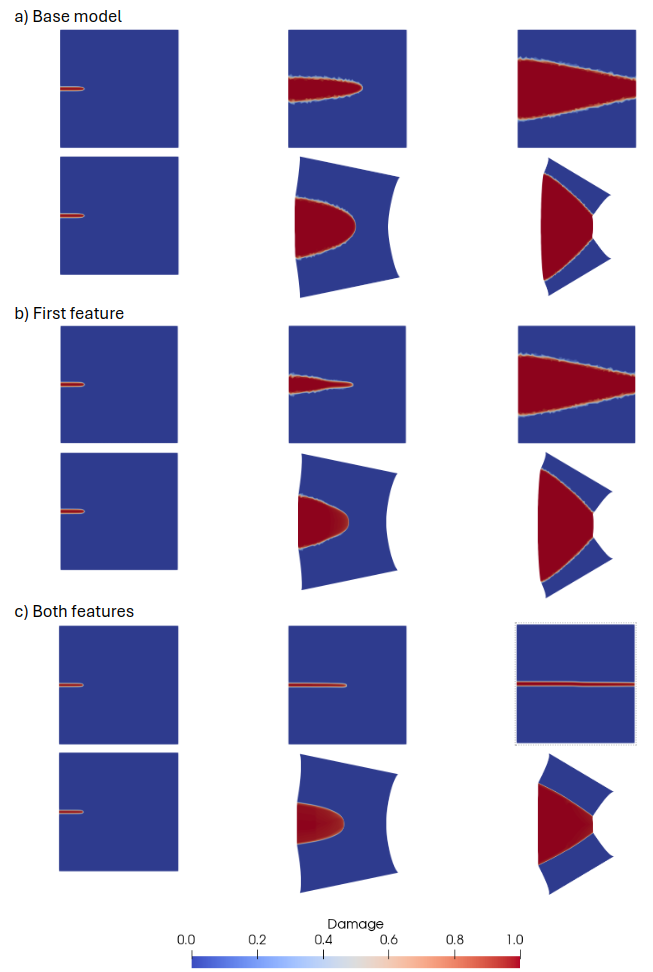}
    \caption{The effect of adding artificial viscosity with $\eta=2$ on damage contours for the three GED modeling scenarios at steps $1$, $35$, and $99$ of the crack propagation (from left to right, respectively).}
    \label{fig:both_constraint_eta2_contours}
\end{figure}

\begin{figure}[h!]
    \centering
    \includegraphics[width=0.8\linewidth]{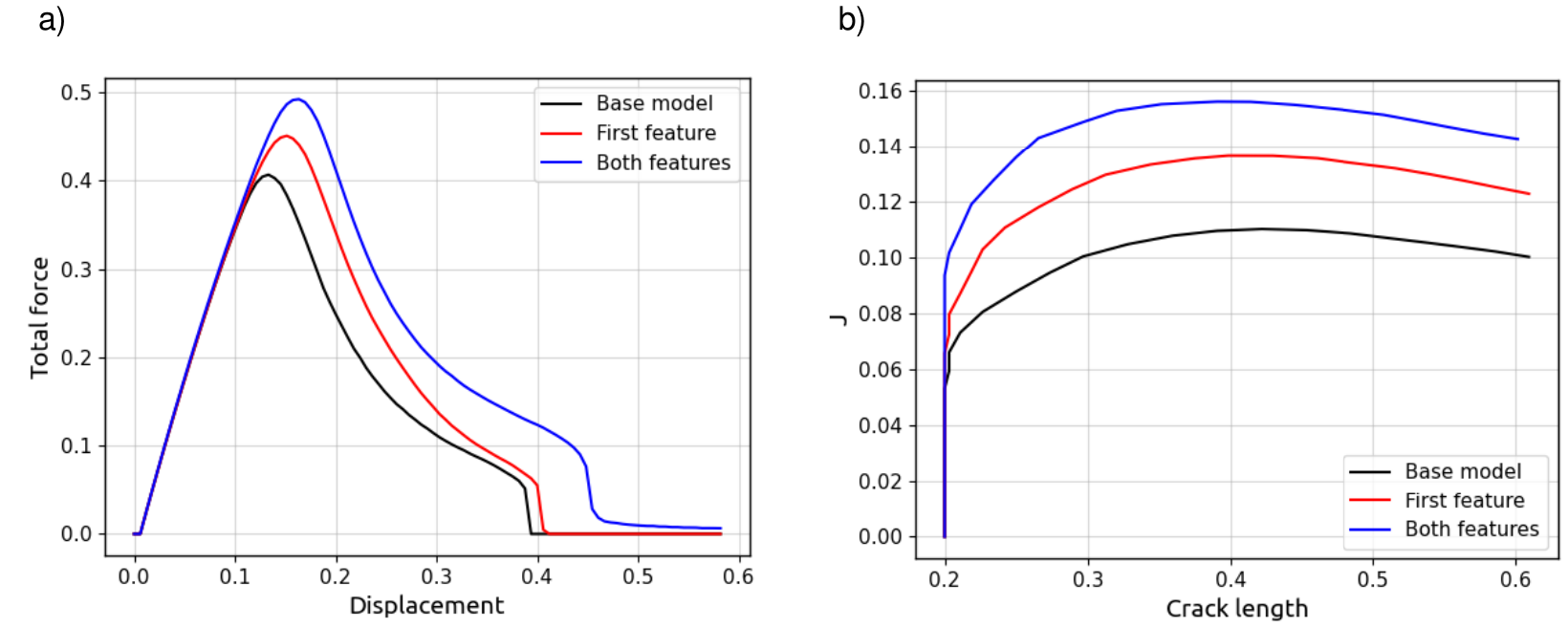}
    \caption{The effect of adding viscosity for the three GED modeling scenarios on the a) force-displacement curve and b) energy release rate. ``Base model'' refers to the model $g(d)=1$ and unbounded nonlocal driving force. ``First feature'' refers to the model with $g(d)=1$ and bounded nonlocal driving force. ``Both features'' refers to the model with decaying $g(d)$ and bounded nonlocal driving force.}
    \label{fig:both_constraint_eta2_graphs}
\end{figure}

To assess the impact of artificial viscosity on the simulation, we developed three GED modeling scenarios based on the examples from Sections \ref{example1}, \ref{example2}, and \ref{example3}, respectively: (i) $g(d)=1$ with an unbounded nonlocal driving force, (ii) $g(d)=1$ with a bounded nonlocal driving force (via Eq. \eqref{GED_strong_viscosity}), and (iii) an activated relaxation function $g(d)$ (via Eq. \eqref{g}) with a bounded nonlocal driving force. The material properties remain the same as those used in the main text, and we here additionally set the artificial viscosity parameter to $\eta = 2$. The artificial viscosity-augmented simulation results are presented in Figs. \ref{fig:both_constraint_eta2_contours} and \ref{fig:both_constraint_eta2_graphs}. Since the artificial viscosity introduces additional dissipation capacity, the material is able to absorb more energy and enhance its resistance to damage and fracture. In this way, crack propagation in the material slows down. This effect is more noticeable when comparing the force-displacement curves in Fig. \ref{fig:both_constraint_eta2_graphs} with Fig. \ref{fig:both_constraint_eta0_graphs}, where the peak of the curve corresponding to the simulation that utilizes both modeling features ($g(d)$ and bounded nonlocal driving force) has increased from $0.45$ to $0.495$ with the incorporation of artificial viscosity, representing a $10\%$ increase. Artificial viscosity further impacts the J-integral curves in Fig. \ref{fig:both_constraint_eta2_graphs} in two ways. First, adding artificial viscosity results in an increased energy release rate during crack propagation. Thus, the J-integral curves in Fig. \ref{fig:both_constraint_eta2_graphs} are generally higher than the corresponding artificial viscosity-free curves in Fig. \ref{fig:both_constraint_eta0_graphs}. Second, the J-integral curves no longer plateau in a straight line. Instead, they curve, as can be seen in Fig. \ref{fig:both_constraint_eta2_graphs}, indicating the dynamic nature of the artificial viscosity effect. These findings align with those reported in \cite{mousavi2024evaluating}.


\newpage
\bibliographystyle{elsarticle-num}

\begin{thebibliography}{100}
\expandafter\ifx\csname url\endcsname\relax
  \def\url#1{\texttt{#1}}\fi
\expandafter\ifx\csname urlprefix\endcsname\relax\def\urlprefix{URL }\fi
\expandafter\ifx\csname href\endcsname\relax
  \def\href#1#2{#2} \def\path#1{#1}\fi

\bibitem{drury2003hydrogels}
J.~L. Drury, D.~J. Mooney, Hydrogels for tissue engineering: scaffold design variables and applications, Biomaterials 24~(24) (2003) 4337--4351.

\bibitem{nonoyama2016double}
T.~Nonoyama, S.~Wada, R.~Kiyama, N.~Kitamura, M.~T.~I. Mredha, X.~Zhang, T.~Kurokawa, T.~Nakajima, Y.~Takagi, K.~Yasuda, et~al., Double-network hydrogels strongly bondable to bones by spontaneous osteogenesis penetration, Advanced Materials 28~(31) (2016) 6740--6745.

\bibitem{bouklas2012swelling}
N.~Bouklas, R.~Huang, Swelling kinetics of polymer gels: comparison of linear and nonlinear theories, Soft Matter 8~(31) (2012) 8194--8203.

\bibitem{bouklas2015effect}
N.~Bouklas, C.~M. Landis, R.~Huang, Effect of solvent diffusion on crack-tip fields and driving force for fracture of hydrogels, Journal of Applied Mechanics 82~(8) (2015).

\bibitem{kang2010variational}
M.~K. Kang, R.~Huang, A variational approach and finite element implementation for swelling of polymeric hydrogels under geometric constraints (2010).

\bibitem{bouklas2015nonlinear}
N.~Bouklas, C.~M. Landis, R.~Huang, A nonlinear, transient finite element method for coupled solvent diffusion and large deformation of hydrogels, Journal of the Mechanics and Physics of Solids 79 (2015) 21--43.

\bibitem{gent2012engineering}
A.~N. Gent, Engineering with {R}ubber: {H}ow to {D}esign {R}ubber {C}omponents, Carl Hanser Verlag GmbH Co KG, 2012.

\bibitem{zhalmuratova2020reinforced}
D.~Zhalmuratova, H.-J. Chung, Reinforced {G}els and {E}lastomers for {B}iomedical and {S}oft {R}obotics {A}pplications, ACS Applied Polymer Materials 2~(3) (2020) 1073--1091.

\bibitem{mark2003elastomers}
J.~Mark, Elastomers with multimodal distributions of network chain lengths, in: Macromolecular Symposia, Vol. 191, Wiley Online Library, 2003, pp. 121--130.

\bibitem{sun2012highly}
J.-Y. Sun, X.~Zhao, W.~R. Illeperuma, O.~Chaudhuri, K.~H. Oh, D.~J. Mooney, J.~J. Vlassak, Z.~Suo, Highly stretchable and tough hydrogels, Nature 489~(7414) (2012) 133--136.

\bibitem{itskov2016rubber}
M.~Itskov, A.~Knyazeva, A rubber elasticity and softening model based on chain length statistics, International Journal of Solids and Structures 80 (2016) 512--519.

\bibitem{tehrani2017effect}
M.~Tehrani, A.~Sarvestani, Effect of chain length distribution on mechanical behavior of polymeric networks, European Polymer Journal 87 (2017) 136--146.

\bibitem{bai2017fatigue}
R.~Bai, Q.~Yang, J.~Tang, X.~P. Morelle, J.~Vlassak, Z.~Suo, Fatigue fracture of tough hydrogels, Extreme Mechanics Letters 15 (2017) 91--96.

\bibitem{chen2017flaw}
C.~Chen, Z.~Wang, Z.~Suo, Flaw sensitivity of highly stretchable materials, Extreme Mechanics Letters 10 (2017) 50--57.

\bibitem{zhou2021flaw}
Y.~Zhou, J.~Hu, P.~Zhao, W.~Zhang, Z.~Suo, T.~Lu, Flaw-sensitivity of a tough hydrogel under monotonic and cyclic loads, Journal of the Mechanics and Physics of Solids 153 (2021) 104483.

\bibitem{yang2019polyacrylamide}
C.~Yang, T.~Yin, Z.~Suo, Polyacrylamide hydrogels. i. network imperfection, Journal of the Mechanics and Physics of Solids 131 (2019) 43--55.

\bibitem{mulderrig2023statistical}
J.~Mulderrig, B.~Talamini, N.~Bouklas, A statistical mechanics framework for polymer chain scission, based on the concepts of distorted bond potential and asymptotic matching, Journal of the Mechanics and Physics of Solids 174 (2023) 105244.

\bibitem{yang2020multiscale}
T.~Yang, K.~M. Liechti, R.~Huang, A multiscale cohesive zone model for rate-dependent fracture of interfaces, Journal of the Mechanics and Physics of Solids 145 (2020) 104142.

\bibitem{slootman2020quantifying}
J.~Slootman, V.~Waltz, C.~J. Yeh, C.~Baumann, R.~G{\"o}stl, J.~Comtet, C.~Creton, Quantifying rate-and temperature-dependent molecular damage in elastomer fracture, Physical Review X 10~(4) (2020) 041045.

\bibitem{verron2017equal}
E.~Verron, A.~Gros, An equal force theory for network models of soft materials with arbitrary molecular weight distribution, Journal of the Mechanics and Physics of Solids 106 (2017) 176--190.

\bibitem{mulderrig2021affine}
J.~Mulderrig, B.~Li, N.~Bouklas, Affine and non-affine microsphere models for chain scission in polydisperse elastomer networks, Mechanics of Materials 160 (2021) 103857.

\bibitem{li2020variational}
B.~Li, N.~Bouklas, A variational phase-field model for brittle fracture in polydisperse elastomer networks, International Journal of Solids and Structures 182 (2020) 193--204.

\bibitem{lamont2021rate}
S.~C. Lamont, J.~Mulderrig, N.~Bouklas, F.~J. Vernerey, Rate-{D}ependent {D}amage {M}echanics of {P}olymer {N}etworks with {R}eversible {B}onds, Macromolecules 54~(23) (2021) 10801--10813.

\bibitem{lin2020fracture}
S.~Lin, X.~Zhao, Fracture of polymer networks with diverse topological defects, Physical Review E 102~(5) (2020) 052503.

\bibitem{lin2021fracture}
S.~Lin, J.~Ni, D.~Zheng, X.~Zhao, Fracture and fatigue of ideal polymer networks, Extreme Mechanics Letters 48 (2021) 101399.

\bibitem{zheng2022fracture}
D.~Zheng, S.~Lin, J.~Ni, X.~Zhao, Fracture and fatigue of entangled and unentangled polymer networks, Extreme Mechanics Letters 51 (2022) 101608.

\bibitem{lei2022network}
J.~Lei, Z.~Liu, A network mechanics method to study the mechanism of the large-deformation fracture of elastomers, Journal of Applied Physics 132~(13) (2022).

\bibitem{ducrot2014toughening}
E.~Ducrot, Y.~Chen, M.~Bulters, R.~P. Sijbesma, C.~Creton, Toughening {E}lastomers with {S}acrificial {B}onds and {W}atching {T}hem {B}reak, Science 344~(6180) (2014) 186--189.

\bibitem{morelle20213d}
X.~P. Morelle, G.~E. Sanoja, S.~Castagnet, C.~Creton, 3{D} fluorescent mapping of invisible molecular damage after cavitation in hydrogen exposed elastomers, Soft Matter 17~(16) (2021) 4266--4274.

\bibitem{sanoja2021mechanical}
G.~E. Sanoja, X.~P. Morelle, J.~Comtet, C.~J. Yeh, M.~Ciccotti, C.~Creton, Why is mechanical fatigue different from toughness in elastomers? {T}he role of damage by polymer chain scission, Science Advances 7~(42) (2021) eabg9410.

\bibitem{slootman2022molecular}
J.~Slootman, C.~J. Yeh, P.~Millereau, J.~Comtet, C.~Creton, A molecular interpretation of the toughness of multiple network elastomers at high temperature, Proceedings of the National Academy of Sciences 119~(13) (2022) e2116127119.

\bibitem{ju2024role}
J.~Ju, G.~E. Sanoja, L.~Cipelletti, M.~Ciccotti, B.~Zhu, T.~Narita, C.~Yuen~Hui, C.~Creton, Role of molecular damage in crack initiation mechanisms of tough elastomers, Proceedings of the National Academy of Sciences 121~(45) (2024) e2410515121.

\bibitem{xiang2018general}
Y.~Xiang, D.~Zhong, P.~Wang, G.~Mao, H.~Yu, S.~Qu, A general constitutive model of soft elastomers, Journal of the Mechanics and Physics of Solids 117 (2018) 110--122.

\bibitem{chen2020mechanically}
S.~Chen, L.~Sun, X.~Zhou, Y.~Guo, J.~Song, S.~Qian, Z.~Liu, Q.~Guan, E.~Meade~Jeffries, W.~Liu, et~al., Mechanically and biologically skin-like elastomers for bio-integrated electronics, Nature communications 11~(1) (2020) 1107.

\bibitem{lake1967strength}
G.~Lake, A.~Thomas, The strength of highly elastic materials, Proceedings of the Royal Society of London. Series A. Mathematical and Physical Sciences 300~(1460) (1967) 108--119.

\bibitem{wang2019quantitative}
S.~Wang, S.~Panyukov, M.~Rubinstein, S.~L. Craig, Quantitative adjustment to the molecular energy parameter in the lake--thomas theory of polymer fracture energy, Macromolecules 52~(7) (2019) 2772--2777.

\bibitem{wang2023contribution}
S.~Wang, S.~Panyukov, S.~L. Craig, M.~Rubinstein, Contribution of unbroken strands to the fracture of polymer networks, Macromolecules 56~(6) (2023) 2309--2318.

\bibitem{beech2023reactivity}
H.~K. Beech, S.~Wang, D.~Sen, D.~Rota, T.~B. Kouznetsova, A.~Arora, M.~Rubinstein, S.~L. Craig, B.~D. Olsen, Reactivity-guided depercolation processes determine fracture behavior in end-linked polymer networks, ACS Macro Letters 12~(12) (2023) 1685--1691.

\bibitem{deng2023nonlocal}
B.~Deng, S.~Wang, C.~Hartquist, X.~Zhao, Nonlocal intrinsic fracture energy of polymerlike networks, Physical Review Letters 131~(22) (2023) 228102.

\bibitem{wang2024loop}
S.~Wang, C.~M. Hartquist, B.~Deng, X.~Zhao, A loop-opening model for the intrinsic fracture energy of polymer networks, Macromolecules (2024).

\bibitem{hartquist2025scaling}
C.~Hartquist, S.~Wang, Q.~Cui, W.~Matusik, B.~Deng, X.~Zhao, Scaling law for intrinsic fracture energy of diverse stretchable networks, Physical Review X 15~(1) (2025) 011002.

\bibitem{hartquist2025fracture}
C.~M. Hartquist, S.~Wang, B.~Deng, H.~K. Beech, S.~L. Craig, B.~D. Olsen, M.~Rubinstein, X.~Zhao, Fracture of polymer-like networks with hybrid bond strengths, Journal of the Mechanics and Physics of Solids 195 (2025) 105931.

\bibitem{arora2020fracture}
A.~Arora, T.-S. Lin, H.~K. Beech, H.~Mochigase, R.~Wang, B.~D. Olsen, Fracture of polymer networks containing topological defects, Macromolecules 53~(17) (2020) 7346--7355.

\bibitem{arora2021coarse}
A.~Arora, T.-S. Lin, B.~D. Olsen, Coarse-grained simulations for fracture of polymer networks: Stress versus topological inhomogeneities, Macromolecules 55~(1) (2021) 4--14.

\bibitem{arora2024effect}
A.~Arora, Effect of spatial heterogeneity on the elasticity and fracture of polymer networks, Macromolecules (2024).

\bibitem{wang2024fracture}
S.-Q. Wang, Z.~Fan, C.~Gupta, A.~Siavoshani, T.~Smith, Fracture behavior of polymers in plastic and elastomeric states, Macromolecules 57~(9) (2024) 3875--3900.

\bibitem{fan2024discontinuous}
Z.~Fan, S.-Q. Wang, Discontinuous dynamic transition in crack propagation of stretchable elastomers, Physical Review E 110~(6) (2024) 065001.

\bibitem{wang2024fresh}
S.-Q. Wang, Z.~Fan, A.~Siavoshani, M.-c. Wang, J.~Wang, Fresh considerations regarding time-dependent elastomeric fracture, Extreme Mechanics Letters (2024) 102277.

\bibitem{persson2024influence}
B.~Persson, Influence of temperature and crack-tip speed on crack propagation in elastic solids, The Journal of Chemical Physics 161~(18) (2024).

\bibitem{yu2025shortest}
Z.~Yu, N.~E. Jackson, Shortest paths govern bond rupture in thermoset networks, Macromolecules (2025).

\bibitem{barney2022fracture}
C.~W. Barney, Z.~Ye, I.~Sacligil, K.~R. McLeod, H.~Zhang, G.~N. Tew, R.~A. Riggleman, A.~J. Crosby, Fracture of model end-linked networks, Proceedings of the National Academy of Sciences 119~(7) (2022) e2112389119.

\bibitem{zhang2024predicting}
H.~Zhang, R.~A. Riggleman, Predicting failure locations in model end-linked polymer networks, Physical Review Materials 8~(3) (2024) 035604.

\bibitem{li2020elongation}
Z.~Li, Z.~Liu, The elongation-criterion for fracture toughness of hydrogels based on percolation model, Journal of Applied Physics 127~(21) (2020).

\bibitem{de2016gradient}
R.~de~Borst, C.~V. Verhoosel, Gradient damage vs phase-field approaches for fracture: Similarities and differences, Computer Methods in Applied Mechanics and Engineering 312 (2016) 78--94.

\bibitem{CAMACHO19962899}
G.~Camacho, M.~Ortiz, \href{https://www.sciencedirect.com/science/article/pii/0020768395002553}{Computational modelling of impact damage in brittle materials}, International Journal of Solids and Structures 33~(20) (1996) 2899--2938.
\newblock \href {https://doi.org/https://doi.org/10.1016/0020-7683(95)00255-3} {\path{doi:https://doi.org/10.1016/0020-7683(95)00255-3}}.
\newline\urlprefix\url{https://www.sciencedirect.com/science/article/pii/0020768395002553}

\bibitem{Ingraffea1985}
A.~R. Ingraffea, V.~Saouma, \href{https://doi.org/10.1007/978-94-009-6152-4_4}{Numerical modeling of discrete crack propagation in reinforced and plain concrete}, Springer Netherlands, Dordrecht, 1985, pp. 171--225.
\newblock \href {https://doi.org/10.1007/978-94-009-6152-4_4} {\path{doi:10.1007/978-94-009-6152-4_4}}.
\newline\urlprefix\url{https://doi.org/10.1007/978-94-009-6152-4_4}

\bibitem{moes1999finite}
N.~Mo{\"e}s, J.~Dolbow, T.~Belytschko, A finite element method for crack growth without remeshing, International journal for numerical methods in engineering 46~(1) (1999) 131--150.

\bibitem{sukumar2000extended}
N.~Sukumar, N.~Mo{\"e}s, B.~Moran, T.~Belytschko, Extended finite element method for three-dimensional crack modelling, International journal for numerical methods in engineering 48~(11) (2000) 1549--1570.

\bibitem{khoei2023modeling}
A.~Khoei, S.~Mousavi, N.~Hosseini, Modeling density-driven flow and solute transport in heterogeneous reservoirs with micro/macro fractures, Advances in Water Resources 182 (2023) 104571.

\bibitem{cazes2009thermodynamic}
F.~Cazes, M.~Coret, A.~Combescure, A.~Gravouil, A thermodynamic method for the construction of a cohesive law from a nonlocal damage model, International Journal of Solids and Structures 46~(6) (2009) 1476--1490.

\bibitem{cuvilliez2012finite}
S.~Cuvilliez, F.~Feyel, E.~Lorentz, S.~Michel-Ponnelle, A finite element approach coupling a continuous gradient damage model and a cohesive zone model within the framework of quasi-brittle failure, Computer methods in applied mechanics and engineering 237 (2012) 244--259.

\bibitem{buche2021chain}
M.~R. Buche, M.~N. Silberstein, Chain breaking in the statistical mechanical constitutive theory of polymer networks, Journal of the Mechanics and Physics of Solids 156 (2021) 104593.

\bibitem{lei2021recent}
J.~Lei, Z.~Li, S.~Xu, Z.~Liu, Recent advances of hydrogel network models for studies on mechanical behaviors, Acta Mechanica Sinica 37 (2021) 367--386.

\bibitem{mao2017rupture}
Y.~Mao, B.~Talamini, L.~Anand, Rupture of polymers by chain scission, Extreme Mechanics Letters 13 (2017) 17--24.

\bibitem{xiao2021modeling}
R.~Xiao, N.~Han, D.~Zhong, S.~Qu, Modeling the mechanical behaviors of multiple network elastomers, Mechanics of Materials 161 (2021) 103992.

\bibitem{zhao2021multiscale}
Z.~Zhao, H.~Lei, H.-S. Chen, Q.~Zhang, P.~Wang, M.~Lei, A multiscale tensile failure model for double network elastomer composites, Mechanics of Materials 163 (2021) 104074.

\bibitem{lu2020pseudo}
T.~Lu, Z.~Wang, J.~Tang, W.~Zhang, T.~Wang, A pseudo-elasticity theory to model the strain-softening behavior of tough hydrogels, Journal of the Mechanics and Physics of Solids 137 (2020) 103832.

\bibitem{arunachala2021energy}
P.~K. Arunachala, R.~Rastak, C.~Linder, Energy based fracture initiation criterion for strain-crystallizing rubber-like materials with pre-existing cracks, Journal of the Mechanics and Physics of Solids 157 (2021) 104617.

\bibitem{dal2009micro}
H.~Dal, M.~Kaliske, A micro-continuum-mechanical material model for failure of rubber-like materials: {A}pplication to ageing-induced fracturing, Journal of the Mechanics and Physics of Solids 57~(8) (2009) 1340--1356.

\bibitem{guo2021micromechanics}
Q.~Guo, F.~Za{\"\i}ri, A micromechanics-based model for deformation-induced damage and failure in elastomeric media, International Journal of Plasticity 140 (2021) 102976.

\bibitem{miehe2010rate}
C.~Miehe, M.~Hofacker, F.~Welschinger, A phase field model for rate-independent crack propagation: Robust algorithmic implementation based on operator splits, Computer Methods in Applied Mechanics and Engineering 199 (2010) 2765--2778.

\bibitem{borden2012phase}
M.~J. Borden, C.~V. Verhoosel, M.~A. Scott, T.~J. Hughes, C.~M. Landis, A phase-field description of dynamic brittle fracture, Computer Methods in Applied Mechanics and Engineering 217 (2012) 77--95.

\bibitem{ambati2015review}
M.~Ambati, T.~Gerasimov, L.~De~Lorenzis, A review on phase-field models of brittle fracture and a new fast hybrid formulation, Computational Mechanics 55~(2) (2015) 383--405.

\bibitem{yin2020ductile}
B.~Yin, M.~Kaliske, A ductile phase-field model based on degrading the fracture toughness: Theory and implementation at small strain, Computer Methods in Applied Mechanics and Engineering 366 (2020) 113068.

\bibitem{kumar2018fracture}
A.~Kumar, G.~A. Francfort, O.~Lopez-Pamies, Fracture and healing of elastomers: A phase-transition theory and numerical implementation, Journal of the Mechanics and Physics of Solids 112 (2018) 523--551.

\bibitem{kumar2020phase}
A.~Kumar, O.~Lopez-Pamies, The phase-field approach to self-healable fracture of elastomers: A model accounting for fracture nucleation at large, with application to a class of conspicuous experiments, Theoretical and Applied Fracture Mechanics (2020) 102550.

\bibitem{talamini2021attaining}
B.~Talamini, M.~R. Tupek, A.~J. Stershic, T.~Hu, J.~W. Foulk~III, J.~T. Ostien, J.~E. Dolbow, Attaining regularization length insensitivity in phase-field models of ductile failure, Computer Methods in Applied Mechanics and Engineering 384 (2021) 113936.

\bibitem{vassilevski1996preconditioning}
P.~S. Vassilevski, R.~D. Lazarov, Preconditioning mixed finite element saddle-point elliptic problems, Numerical linear algebra with applications 3~(1) (1996) 1--20.

\bibitem{benzi2005numerical}
M.~Benzi, G.~H. Golub, J.~Liesen, Numerical solution of saddle point problems, Acta numerica 14 (2005) 1--137.

\bibitem{loghin2004analysis}
D.~Loghin, A.~J. Wathen, Analysis of preconditioners for saddle-point problems, SIAM Journal on Scientific Computing 25~(6) (2004) 2029--2049.

\bibitem{taylor2000mixed}
R.~L. Taylor, A mixed-enhanced formulation tetrahedral finite elements, International Journal for Numerical Methods in Engineering 47~(1-3) (2000) 205--227.

\bibitem{onate2004finitecalculus}
E.~Oñate, J.~Rojek, R.~L. Taylor, O.~C. Zienkiewicz, Finite calculus formulation for incompressible solids using linear triangles and tetrahedra, International Journal for Numerical Methods in Engineering 59~(11) (2004) 1473--1500.

\bibitem{gavagnin2020stabilized}
C.~Gavagnin, L.~Sanavia, L.~De~Lorenzis, Stabilized mixed formulation for phase-field computation of deviatoric fracture in elastic and poroelastic materials, Computational Mechanics 65~(6) (2020) 1447--1465.

\bibitem{mang2020phase}
K.~Mang, T.~Wick, W.~Wollner, A phase-field model for fractures in nearly incompressible solids, Computational Mechanics 65~(1) (2020) 61--78.

\bibitem{alessi2020phase}
R.~Alessi, F.~Freddi, L.~Mingazzi, Phase-field numerical strategies for deviatoric driven fractures, Computer Methods in Applied Mechanics and Engineering 359 (2020) 112651.

\bibitem{suh2020phase}
H.~S. Suh, W.~Sun, D.~T. O’Connor, A phase field model for cohesive fracture in micropolar continua, Computer Methods in Applied Mechanics and Engineering 369 (2020) 113181.

\bibitem{cajuhi2018phase}
T.~Cajuhi, L.~Sanavia, L.~De~Lorenzis, Phase-field modeling of fracture in variably saturated porous media, Computational Mechanics 61~(3) (2018) 299--318.

\bibitem{wriggers2021taylor}
P.~Wriggers, M.~De~Bellis, B.~Hudobivnik, A taylor--hood type virtual element formulations for large incompressible strains, Computer Methods in Applied Mechanics and Engineering 385 (2021) 114021.

\bibitem{klaas1999stabilized}
O.~Klaas, A.~Maniatty, M.~S. Shephard, A stabilized mixed finite element method for finite elasticity.: Formulation for linear displacement and pressure interpolation, Computer Methods in Applied Mechanics and Engineering 180~(1-2) (1999) 65--79.

\bibitem{maniatty2002higher}
A.~M. Maniatty, Y.~Liu, O.~Klaas, M.~S. Shephard, Higher order stabilized finite element method for hyperelastic finite deformation, Computer Methods in Applied Mechanics and Engineering 191~(13-14) (2002) 1491--1503.

\bibitem{ang2022stabilized}
I.~Ang, N.~Bouklas, B.~Li, Stabilized formulation for phase-field fracture in nearly incompressible hyperelasticity, International Journal for Numerical Methods in Engineering 123~(19) (2022) 4655--4673.

\bibitem{swamynathan2022phase}
S.~Swamynathan, S.~Jobst, D.~Kienle, M.-A. Keip, Phase-field modeling of fracture in strain-hardening elastomers: Variational formulation, multiaxial experiments and validation, Engineering Fracture Mechanics 265 (2022) 108303.

\bibitem{arunachala2023multiscale}
P.~K. Arunachala, S.~A. Vajari, M.~Neuner, C.~Linder, A multiscale phase field fracture approach based on the non-affine microsphere model for rubber-like materials, Computer Methods in Applied Mechanics and Engineering 410 (2023) 115982.

\bibitem{feng2023phase}
H.~Feng, L.~Jiang, Phase field modeling on fracture behaviors of elastomers considering deformation-dependent and damage-dependent material viscosity, Engineering Fracture Mechanics 292 (2023) 109621.

\bibitem{ye2023nonlinear}
J.-Y. Ye, R.~Ballarini, L.-W. Zhang, A nonlinear and rate-dependent fracture phase field framework for multiple cracking of polymer, Computer Methods in Applied Mechanics and Engineering 410 (2023) 116017.

\bibitem{zhao2023phase}
Z.~Zhao, P.~Wang, S.~Duan, M.~Lei, H.~Lei, A phase field model for the damage and fracture of multiple network elastomers, Journal of Applied Mechanics 90~(2) (2023) 021006.

\bibitem{pranavi2024unifying}
D.~Pranavi, P.~Steinmann, A.~Rajagopal, A unifying finite strain modeling framework for anisotropic mixed-mode fracture in soft materials, Computational Mechanics 73~(1) (2024) 123--137.

\bibitem{kuhn1942beziehungen}
W.~Kuhn, F.~Gr{\"u}n, Beziehungen zwischen elastischen {K}onstanten und {D}ehnungsdoppelbrechung hochelastischer {S}toffe, Kolloid-Zeitschrift 101 (1942) 248--271.

\bibitem{smith1996overstretching}
S.~B. Smith, Y.~Cui, C.~Bustamante, Overstretching {B}-{DNA}: {T}he {E}lastic {R}esponse of {I}ndividual {D}ouble-{S}tranded and {S}ingle-{S}tranded {DNA} {M}olecules, Science 271~(5250) (1996) 795--799.

\bibitem{buche2022freely}
M.~R. Buche, M.~N. Silberstein, S.~J. Grutzik, Freely jointed chain models with extensible links, Physical Review E 106~(2) (2022) 024502.

\bibitem{diani2019fully}
J.~Diani, P.~Le~Tallec, A fully equilibrated microsphere model with damage for rubberlike materials, Journal of the Mechanics and Physics of Solids 124 (2019) 702--713.

\bibitem{wang1952statistical}
M.~C. Wang, E.~Guth, Statistical {T}heory of {N}etworks of {N}on-{G}aussian {F}lexible {C}hains, The Journal of Chemical Physics 20~(7) (1952) 1144--1157.

\bibitem{flory1943statistical}
P.~J. Flory, J.~Rehner~Jr, Statistical {M}echanics of {C}ross-{L}inked {P}olymer {N}etworks {I}. {R}ubberlike {E}lasticity, The Journal of Chemical Physics 11~(11) (1943) 512--520.

\bibitem{arruda1993three}
E.~M. Arruda, M.~C. Boyce, A three-dimensional constitutive model for the large stretch behavior of rubber elastic materials, Journal of the Mechanics and Physics of Solids 41~(2) (1993) 389--412.

\bibitem{treloar1979non}
L.~Treloar, G.~Riding, A non-{G}aussian theory for rubber in biaxial strain. {I}. {M}echanical properties, Proceedings of the Royal Society of London. A. Mathematical and Physical Sciences 369~(1737) (1979) 261--280.

\bibitem{wu1992improved}
P.~Wu, E.~van~der Giessen, On {I}mproved 3-{D} {N}on-{G}aussian {N}etwork {M}odels for {R}ubber {E}lasticity, Mechanics Research Communications 19~(5) (1992) 427--433.

\bibitem{wu1993improved}
P.~Wu, E.~Van Der~Giessen, On improved network models for rubber elasticity and their applications to orientation hardening in glassy polymers, Journal of the Mechanics and Physics of Solids 41~(3) (1993) 427--456.

\bibitem{miehe2004micro}
C.~Miehe, S.~G{\"o}ktepe, F.~Lulei, A micro-macro approach to rubber-like materials—{P}art {I}: the non-affine micro-sphere model of rubber elasticity, Journal of the Mechanics and Physics of Solids 52~(11) (2004) 2617--2660.

\bibitem{tkachuk2012maximal}
M.~Tkachuk, C.~Linder, The maximal advance path constraint for the homogenization of materials with random network microstructure, Philosophical Magazine 92~(22) (2012) 2779--2808.

\bibitem{ghaderi2020physics}
A.~Ghaderi, V.~Morovati, R.~Dargazany, A {P}hysics-{I}nformed {A}ssembly of {F}eed-{F}orward {N}eural {N}etwork {E}ngines to {P}redict {I}nelasticity in {C}ross-{L}inked {P}olymers, Polymers 12~(11) (2020) 2628.

\bibitem{rastak2018non}
R.~Rastak, C.~Linder, A non-affine micro-macro approach to strain-crystallizing rubber-like materials, Journal of the Mechanics and Physics of Solids 111 (2018) 67--99.

\bibitem{govindjee2019fully}
S.~Govindjee, M.~J. Zoller, K.~Hackl, A fully-relaxed variationally-consistent framework for inelastic micro-sphere models: {F}inite viscoelasticity, Journal of the Mechanics and Physics of Solids 127 (2019) 1--19.

\bibitem{xiao2021micromechanical}
R.~Xiao, T.-T. Mai, K.~Urayama, J.~P. Gong, S.~Qu, Micromechanical modeling of the multi-axial deformation behavior in double network hydrogels, International Journal of Plasticity 137 (2021) 102901.

\bibitem{zhan2023new}
L.~Zhan, S.~Wang, S.~Qu, P.~Steinmann, R.~Xiao, A new micro--macro transition for hyperelastic materials, Journal of the Mechanics and Physics of Solids 171 (2023) 105156.

\bibitem{peerlings1996gradient}
R.~H. Peerlings, R.~de~Borst, W.~M. Brekelmans, J.~de~Vree, Gradient enhanced damage for quasi-brittle materials, International Journal for numerical methods in engineering 39~(19) (1996) 3391--3403.

\bibitem{pham2010gradient}
K.~Pham, H.~Amor, J.-J. Marigo, C.~Maurini, Gradient damage models and their use to approximate brittle fracture, International Journal of Damage Mechanics 20~(4) (2011) 618--652.

\bibitem{kuhl2000anisotropic}
E.~Kuhl, E.~Ramm, R.~de~Borst, An anisotropic gradient damage model for quasi-brittle materials, Computer Methods in Applied Mechanics and Engineering 183~(1-2) (2000) 87--103.

\bibitem{marigo2016overview}
J.-J. Marigo, C.~Maurini, K.~Pham, An overview of the modelling of fracture by gradient damage models, Meccanica 51~(12) (2016) 3107--3128.

\bibitem{seupel2019gradient}
A.~Seupel, M.~Kuna, A gradient-enhanced damage model motivated by engineering approaches to ductile failure of steels, International Journal of Damage Mechanics 28~(8) (2019) 1261--1296.

\bibitem{zhao2023modified}
D.~Zhao, B.~Yin, S.~Tarachandani, M.~Kaliske, A modified cap plasticity description coupled with a localizing gradient-enhanced approach for concrete failure modeling, Computational Mechanics 72~(4) (2023) 787--801.

\bibitem{saji2024new}
R.~P. Saji, P.~Pantidis, M.~E. Mobasher, A new unified arc-length method for damage mechanics problems, Computational Mechanics (2024) 1--32.

\bibitem{peerlings1998gradient}
R.~H. Peerlings, R.~de~Borst, W.~Brekelmans, M.~G. Geers, Gradient-enhanced damage modelling of concrete fracture, Mechanics of Cohesive-frictional Materials: An International Journal on Experiments, Modelling and Computation of Materials and Structures 3~(4) (1998) 323--342.

\bibitem{lorentz2011gradient}
E.~Lorentz, V.~Godard, Gradient damage models: Toward full-scale computations, Computer Methods in Applied Mechanics and Engineering 200~(21-22) (2011) 1927--1944.

\bibitem{lorentz2012modelling}
E.~Lorentz, S.~Cuvilliez, K.~Kazymyrenko, Modelling large crack propagation: from gradient damage to cohesive zone models, International journal of fracture 178~(1) (2012) 85--95.

\bibitem{lorentz2017nonlocal}
E.~Lorentz, A nonlocal damage model for plain concrete consistent with cohesive fracture, International Journal of Fracture 207~(2) (2017) 123--159.

\bibitem{talamini2018progressive}
B.~Talamini, Y.~Mao, L.~Anand, Progressive damage and rupture in polymers, Journal of the Mechanics and Physics of Solids 111 (2018) 434--457.

\bibitem{valverde2023locking}
A.~Valverde-Gonz{\'a}lez, J.~Reinoso, B.~Dortdivanlioglu, M.~Paggi, Locking treatment of penalty-based gradient-enhanced damage formulation for failure of compressible and nearly incompressible hyperelastic materials, Computational Mechanics 72~(4) (2023) 635--662.

\bibitem{lamm2024gradient1}
L.~Lamm, J.~Pfeifer, H.~Holthusen, B.~Schaaf, R.~Seewald, A.~Schiebahn, T.~Brepols, M.~Feldmann, U.~Reisgen, S.~Reese, Gradient-extended damage modelling for polymeric materials at finite strains: Rate-dependent damage evolution combined with viscoelasticity, European Journal of Mechanics-A/Solids 103 (2024) 105121.

\bibitem{forest2009micromorphic}
S.~Forest, Micromorphic approach for gradient elasticity, viscoplasticity, and damage, Journal of Engineering Mechanics 135~(3) (2009) 117--131.

\bibitem{lamm2024gradient2}
L.~Lamm, A.~Awad, J.~Pfeifer, H.~Holthusen, S.~Felder, S.~Reese, T.~Brepols, A gradient-extended thermomechanical model for rate-dependent damage and failure within rubberlike polymeric materials at finite strains, International Journal of Plasticity 173 (2024) 103883.

\bibitem{mousavi2024evaluating}
S.~M. Mousavi, I.~Ang, J.~Mulderrig, N.~Bouklas, Evaluating fracture energy predictions using phase-field and gradient-enhanced damage models for elastomers, Journal of Applied Mechanics (2024) 1--14.

\bibitem{wosatko2021comparison}
A.~Wosatko, Comparison of evolving gradient damage formulations with different activity functions, Archive of Applied Mechanics 91~(2) (2021) 597--627.

\bibitem{wosatko2022survey}
A.~Wosatko, {S}urvey of {L}ocalizing {G}radient {D}amage in {S}tatic and {D}ynamic {T}ension of {C}oncrete, Materials 15~(5) (2022) 1875.

\bibitem{alnaes2015fenics}
M.~Aln{\ae}s, J.~Blechta, J.~Hake, A.~Johansson, B.~Kehlet, A.~Logg, C.~Richardson, J.~Ring, M.~E. Rognes, G.~N. Wells, The fenics project version 1.5, Archive of Numerical Software 3~(100) (2015).

\bibitem{holzapfel2002nonlinear}
G.~A. Holzapfel, Nonlinear solid mechanics: a continuum approach for engineering science (2002).

\bibitem{Arruda-Boyce1993}
E.~M. Arruda, M.~C. Boyce, A three-dimensional constitutive model for the large stretch behavior of rubber elastic materials, Journal of the Mechanics and Physics of Solids 41~(2) (1993) 389--412.

\bibitem{wriggers2006computational}
P.~Wriggers, T.~A. Laursen, Computational contact mechanics, Vol.~2, Springer, 2006.

\bibitem{poh2017localizing}
L.~H. Poh, G.~Sun, Localizing gradient damage model with decreasing interactions, International Journal for Numerical Methods in Engineering 110~(6) (2017) 503--522.

\bibitem{sarkar2020thermo}
S.~Sarkar, I.~V. Singh, B.~Mishra, A thermo-mechanical gradient enhanced damage method for fracture, Computational Mechanics 66 (2020) 1399--1426.

\bibitem{geers1998strain}
M.~Geers, R.~De~Borst, W.~Brekelmans, R.~Peerlings, Strain-based transient-gradient damage model for failure analyses, Computer methods in applied mechanics and engineering 160~(1-2) (1998) 133--153.

\bibitem{saroukhani2013simplified}
S.~Saroukhani, R.~Vafadari, A.~Simone, A simplified implementation of a gradient-enhanced damage model with transient length scale effects, Computational Mechanics 51 (2013) 899--909.

\bibitem{sarkar2019comparative}
S.~Sarkar, I.~V. Singh, B.~Mishra, A.~Shedbale, L.~Poh, A comparative study and abaqus implementation of conventional and localizing gradient enhanced damage models, Finite Elements in Analysis and Design 160 (2019) 1--31.

\bibitem{verhoosel2011isogeometric}
C.~V. Verhoosel, M.~A. Scott, T.~J. Hughes, R.~De~Borst, An isogeometric analysis approach to gradient damage models, International Journal for Numerical Methods in Engineering 86~(1) (2011) 115--134.

\bibitem{peerlings2004thermodynamically}
R.~Peerlings, T.~Massart, M.~Geers, A thermodynamically motivated implicit gradient damage framework and its application to brick masonry cracking, Computer Methods in Applied Mechanics and Engineering 193~(30-32) (2004) 3403--3417.

\bibitem{li1985comparison}
F.~Z. Li, C.~F. Shih, A.~Needleman, A comparison of methods for calculating energy release rates, Engineering fracture mechanics 21~(2) (1985) 405--421.

\bibitem{miehe2014phase}
C.~Miehe, L.-M. Sch{\"a}nzel, Phase field modeling of fracture in rubbery polymers. part i: Finite elasticity coupled with brittle failure, Journal of the Mechanics and Physics of Solids 65 (2014) 93--113.

\bibitem{yin2020fracture}
B.~Yin, M.~Kaliske, Fracture simulation of viscoelastic polymers by the phase-field method, Computational Mechanics 65 (2020) 293--309.

\bibitem{najmeddine2024physics}
A.~Najmeddine, M.~Shakiba, Physics and chemistry-based phase-field constitutive framework for thermo-chemically aged elastomer, International Journal of Mechanical Sciences 262 (2024) 108721.

\bibitem{ulloa2022variational}
J.~Ulloa, N.~Noii, R.~Alessi, F.~Aldakheel, G.~Degrande, S.~Fran{\c{c}}ois, Variational modeling of hydromechanical fracture in saturated porous media: A micromechanics-based phase-field approach, Computer Methods in Applied Mechanics and Engineering 396 (2022) 115084.

\bibitem{aldakheel2021multilevel}
F.~Aldakheel, N.~Noii, T.~Wick, O.~Allix, P.~Wriggers, Multilevel global--local techniques for adaptive ductile phase-field fracture, Computer Methods in Applied Mechanics and Engineering 387 (2021) 114175.

\bibitem{miehe2010thermodynamicPF}
C.~Miehe, F.~Welschinger, M.~Hofacker, Thermodynamically consistent phase-field models of fracture: Variational principles and multi-field fe implementations, International Journal for Numerical Methods in Engineering 83~(10) (2010) 1273--1311.

\end{thebibliography}

\end{document}